\documentclass[acmtog,table,nonacm]{acmart}
\usepackage{mymacros}
\usepackage{xspace}
\usepackage{sistyle}
\usepackage{stfloats}
\usepackage{xcolor}
\usepackage[capitalise]{cleveref}
\usepackage{mathtools}
\usepackage{float}
\usepackage{pifont}%
\SIthousandsep{,}
            
\AtBeginDocument{%
  \providecommand\BibTeX{{%
    \normalfont B\kern-0.5em{\scshape i\kern-0.25em b}\kern-0.8em\TeX}}}

\setcopyright{acmlicensed}
\copyrightyear{2024}
\acmYear{2024}
\acmDOI{xxxxxxx.xxxxxxx}

\acmConference[ACM Trans. Graph. Vol. 43, No. 4]{Make sure to enter the correct
  conference title from your rights confirmation emai}{Article 1}{Publication date: July 2024}
\acmBooktitle{ACM Trans. on Graphics}

\acmSubmissionID{739}

\begin{document}

\newcommand{\abbrtitle}{One Noise to Rule Them All}
\title[\abbrtitle]{\abbrtitle:\\
Learning a Unified Model of Spatially-Varying Noise Patterns}

\author{Arman Maesumi}
\email{arman.maesumi@gmail.com}
\orcid{0000-0001-7898-8061}
\affiliation{%
  \institution{Brown University}
  \city{Providence}
  \state{RI}
  \country{USA}
}

\author{Dylan Hu}
\authornote{Both authors contributed equally to this research.}
\email{dh@brown.edu}
\orcid{0009-0001-9456-9470}
\author{Krishi Saripalli}
\authornotemark[1]
\email{krishi_saripalli@brown.edu}
\orcid{0009-0000-5224-2632}
\affiliation{%
  \institution{Brown University}
  \city{Providence}
  \state{RI}
  \country{USA}
}

\author{Vladimir G. Kim}
\email{vova.g.kim@gmail.com}
\orcid{0000-0002-3996-6588}
\affiliation{%
  \institution{Adobe Research}
  \country{USA}
}

\author{Matthew Fisher}
\email{techmatt@gmail.com}
\orcid{0000-0002-8908-3417}
\affiliation{%
  \institution{Adobe Research}
  \country{USA}
}

\author{S\"{o}ren Pirk}
\email{soeren.pirk@gmail.com}
\orcid{0000-0003-1937-9797}
\affiliation{%
  \institution{CAU}
  \city{Kiel}
  \country{Germany}
}

\author{Daniel Ritchie}
\email{daniel_ritchie@brown.edu}
\orcid{0000-0002-8253-0069}
\affiliation{%
  \institution{Brown University}
  \city{Providence}
  \state{RI}
  \country{USA}
}

\renewcommand{\shortauthors}{Maesumi et al.}

\begin{abstract}
Procedural noise is a fundamental component of computer graphics pipelines, offering a flexible way to generate textures  that exhibit ``natural'' random variation. 
Many different types of noise exist, each produced by a separate algorithm.
In this paper, we present a single generative model which can learn to generate multiple types of noise as well as blend between them.
In addition, it is capable of producing spatially-varying noise blends despite not having access to such data for training.
These features are enabled by training a denoising diffusion model using a novel combination of data augmentation and network conditioning techniques.
Like procedural noise generators, the model's behavior is controllable via interpretable parameters plus a source of randomness.
We use our model to produce a variety of visually compelling noise textures.
We also present an application of our model to improving inverse procedural material design; using our model in place of fixed-type noise nodes in a procedural material graph results in higher-fidelity material reconstructions without needing to know the type of noise in advance. Open-sourced materials can be found at \href{https://armanmaesumi.github.io/onenoise/}{\color{cyan}{https://armanmaesumi.github.io/onenoise/}}

\end{abstract}



\begin{teaserfigure}
  \includegraphics[width=\textwidth]{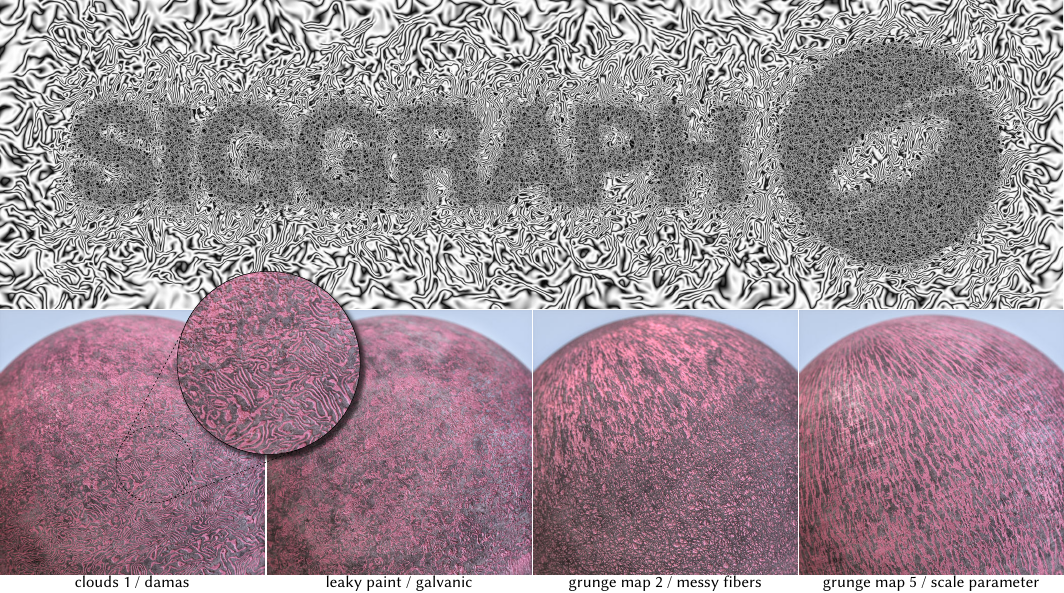}
  \caption{Our method enables the synthesis of a wide range of noise patterns with spatially-varying characteristics. Here we show the flexibility of our unified noise model, allowing one to art-direct their noise in a granular fashion. Our model creates semantically meaningful interpolations between noise configurations; above we see the Siggraph logo written with \texttt{hay fibers} that are nested inside of \texttt{Damascus steel} striations -- the scale and distortion of the steel pattern naturally interpolates into a denser pattern before transitioning into fibers. We also show renderings of a clay shader that incorporates our spatially-varying noise patterns. The first three images make use of class-interpolated noise, the final image uses parameter-interpolated noise. Please zoom into the figures for full visual detail.}
  \label{fig:teaser}
\end{teaserfigure}

\received{20 February 2007}
\received[revised]{12 March 2009}
\received[accepted]{5 June 2009}

\maketitle

\section{Introduction}
Procedural noise has long been a fundamental building block in computer graphics, serving as a versatile tool for modeling fine, naturalistic details in a range of applications. It finds widespread use in creating albedo textures, bump and normal maps, terrain height fields, and density fields for volumetric phenomena such as clouds and smoke. There exists a zoo of algorithms for generating such noise patterns, for instance Perlin, Worley, and Gabor noise \cite{perlin85, worleyNoise, lagae2009gabor}, along with systems for composing and transforming these noise patterns to synthesis more complex shaders (e.g. Blender, Adobe Substance 3D Designer, and other material editors \cite{blender, adobeSubstance}).

\begin{figure}
    \centering
    \includegraphics[width=\linewidth]{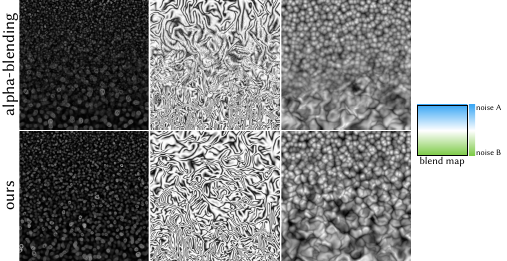}
    \caption{Our model produces noise patterns whose characteristics (i.e. scale, ripples, distortion, etc.) interpolate naturally, creating seamless and coherent transitions. By contrast, traditional alpha-blending results in images with overlapping features, inconsistent feature opacity, and a lack of sensible transitions between the noise characteristics.}
    \label{fig:alphaBlending}
\end{figure}

Despite the advancement of these tools, a fundamental limitation remains in the design process: the necessity to make discrete choices regarding the types of noise to employ. This constraint poses challenges in creative design, where the ideal noise type may not be immediately present among the provided ``zoo'', or where desired visual characteristics fall between the behaviors of existing noise types. Designers might envision patterns exhibiting different noise characteristics in various regions, transitioning smoothly without abrupt or artificial blending. The standard technique for blending between noise types---alpha blending---often yields unsatisfactory results, where features do not naturally interpolate in the image (see Figure \ref{fig:alphaBlending}).

This limitation also complicates inverse design tasks, such as recovering a procedural representation (i.e. a material node graph) of a texture from observed data. 
Such graphs often use multiple noise generator nodes; if the types of these nodes aren't known in advance, the search space for the inverse design problem grows combinatorially, and the need to search over discrete possible noise types limits the applicability of gradient-based optimization.

In this paper, we address these challenges by introducing a method for learning a continuous space of spatially-varying noise patterns from data that lacks any spatially-varying observations. We leverage a denoising diffusion probabilistic model (DDPM) \cite{ho2020ddpm} equipped with a spatially-varying conditioning module, which is trained via a novel application of CutMix data augmentation \cite{yun2019cutmix}. Our method enables the generation of noise that can vary smoothly across the canvas, offering a greater level of control over the resulting noise pattern.
Just as with traditional procedural noise generators, the model can be controlled by changing a set of interpretable parameters or by changing the input source of randomness (i.e. random seed).
Importantly, our method can generate noise images at any requested image size (beyond the size of the training data) and can also produce seamlessly tileable images.

We evaluate our model by using it to generate a variety of  noise blends and spatially-varying noise patterns (including spatial variation driven by image-based masks).
We also present a user interface for painting spatially-varying noise patterns, videos of this interface are included in the supplemental material.
Finally, we show an application of our method to inverse procedural material design, demonstrating that using our model as a noise generator node in a material graph to be optimized can result in higher-fidelity material reconstructions without needing to know the particular noise type for that node in advance.

In summary, our contributions are:
\begin{enumerate}
    \item A generative model which can learn to generate spatially-varying noise patterns.
    \item A training scheme for the model using a novel application of CutMix augmentation, allowing the model to learn to generate spatially-varying noise patterns without having any spatially-varying training data.
    \item An application of our model to inverse procedural material design, showing improved material reconstructions without pre-specified discrete noise types.
\end{enumerate}

\section{Related Work}
\paragraph{Procedural noise}
``Noise is the \textit{random number generator of computer graphics}'' \cite{lagae2010noiseSurvey}. Procedural noise functions collectively form a body of techniques for algorithmically synthesizing patterns that mimic the randomness and irregularities in natural phenomena. These noise functions have become foundational to many applications within computer graphics, facilitating the synthesis of natural materials \cite{perlin85,dorsey2006Patinas}, terrain \cite{genevaux2013terrain,galin2019terrainModelingReview,van2013proceduralDungeons, fournier1982subdivision}, and motion \cite{hinsinger2002interactiveOceans}.

There are a wide variety of noise functions that synthesize distinct patterns, for example, Perlin \cite{perlin85, perlin2002improving}, Wavelet \cite{cook2005wavelet}, and Gabor noise \cite{lagae2009gabor}. Such noises are often composed and transformed using various primitive operations (e.g. scalar functions and domain warping), forming patterns that are colloquially referred to as \emph{noise} in computer graphics --- while these resulting patterns may not strictly fall under the formal definition of noise \cite{musgrave2002texturingBook}, we continue to use the term in this established, albeit technically imprecise sense, throughout the paper. A selection of various noise patterns is shown in Figure \ref{fig:dataset}. For further background, we refer to the survey by Lagae et al.~\shortcite{lagae2010noiseSurvey}.

\paragraph{Neural parametric texture synthesis}\label{sec:relatedWorkTextureSynth}
As noise is a central building block for textures, our work is related to methods for texture synthesis.
Texture synthesis has a deep, decades-long history, a full discussion of which is outside the scope of this paper.
The most relevant methods to our work are parametric texture synthesis methods which use neural networks as their parametric model.
Most of this prior work focuses on \emph{example-based texture synthesis}, i.e. generating textures which are similar in some statistical sense to an input exemplar.
Some methods do this by optimizing an image such that the features produced by passing it through a certain pre-trained neural network match those produced by the exemplar image~\cite{gatys2015texture,DeepCorrelations}.
Others train feedforward neural networks to directly generate texture similar to an input exemplar~\cite{TextureNetworks,SeamlessGAN,MarkovianGAN,TexSynAdversarialExpansion,TexSynGuidedCorrespondence}.
The closest work to ours is a generative adversarial network (GAN) that does not perform example-based synthesis, but rather learns a continuous latent space of textures~\cite{PeriodicSpatialGAN}.
This model supports interpolation between textures, but it is not controllable by a combination of interpretable noise parameters and a source of randomness (its latent space entangles texture ``style'' with spatial randomness).
We initially experimented with a GAN-based architecture and found that it struggled to capture a data distribution with the many different modes represented by a set of highly distinct noise types.
Prior work suggests that diffusion models such as ours fare better at modeling such distributions~\cite{xiao2022tackling}. 

We also acknowledge the role of non-parametric methods in this field. These approaches synthesize textures by using the structures present within an image, without relying on a learned parametric model. For instance, Matusik et al.~\shortcite{simplicialComplex} frame the texture interpolation problem through \emph{morphable models}~\cite{morphableModels}, and define a space of interpolatable textures via a simplicial complex induced by a texture similarity metric. ImageMelding~\cite{ImageMelding}, on the other hand, proposes a patch-based method that smoothly ``in-fills'' regions of a texture via a regularized screened Poisson equation. For a more comprehensive review of this body of work, we refer to the survey by Raad et al.~\shortcite{raad2018textureSurvey}.

\paragraph{Inverse procedural material design}
Recovering procedural representations of materials from observations (i.e. photographs) has become an increasingly promising area of research. Material graphs---the procedural representation of choice---feature nodes that generate, transform, and filter various signals that ultimately become spatially-varying material maps (SVBRDFs). Authoring and editing these material graphs is not only time consuming, but also requires mastery of complex software~\cite{adobeSubstance, blender}. \emph{Inverse procedural material design} aims to alleviate this difficulty by recovering a material graph from a given image. Recent works tackle this problem in various ways, either by 1) directly predicting the parameters of a given material graph~\cite{hu2019InverseProc, guo2020bayesianProcMat}, 2) optimizing the continuous parameters of the graph via gradient-based optimization~\cite{Shi2020MATch, hu2022diffProxy}, or 3) by synthesizing a material graph from scratch~\cite{guerrero2022matformer, zhou2023photomat}.

Most relevant to our work are learned differentiable proxies, which serve as continuous relaxations for gradient-based optimization strategies to the inverse design problem. Hu et al.~\shortcite{hu2022diffProxy} train several GAN-based generative models to act as differentiable proxies for \emph{pattern} generators (i.e. a brick tiling generator node). However, these proxies are deterministic, meaning they cannot model noise functions, and critically, each pattern is represented by a separate generative model. Our unified noise DDPM is able to simultaneously capture many non-deterministic noise functions in the same continuous space, acting as a relaxation over an entire space of unique noise functions.

\begin{figure*}[t]
    \centering
    \includegraphics[width=\linewidth]{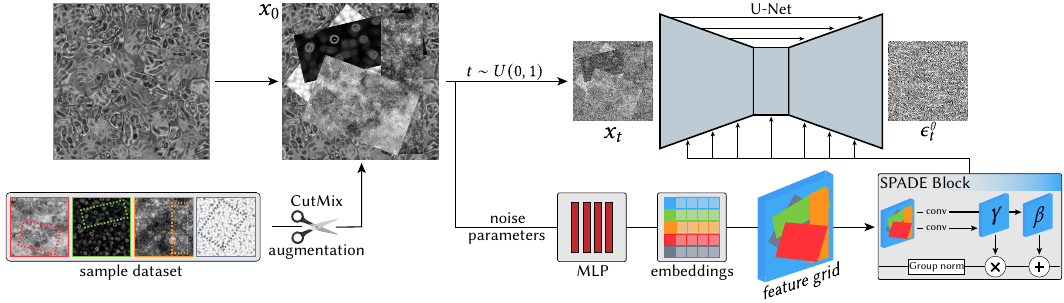}
    \caption{Our DDPM is trained using CutMix data augmentation. We first transform the current data sample (highlighted in blue) by cutting and patching together a set of other random samples from the dataset, resulting in a training image $\*x_0$. The noise parameters for each image patch are passed to an MLP, which projects the parameter sets into an embedding space that encodes both the noise type (class) and the noise parameters. The resulting feature vectors are tiled to form a feature grid, which is used as a conditioning signal in the U-Net's SPADE blocks, as outlined in Section \ref{sec:method_conditioning}.}
    \label{fig:pipeline}
\end{figure*}
\begin{figure*}[b]
    \centering
    \includegraphics[width=\linewidth, trim=0 0.3cm 0 0]{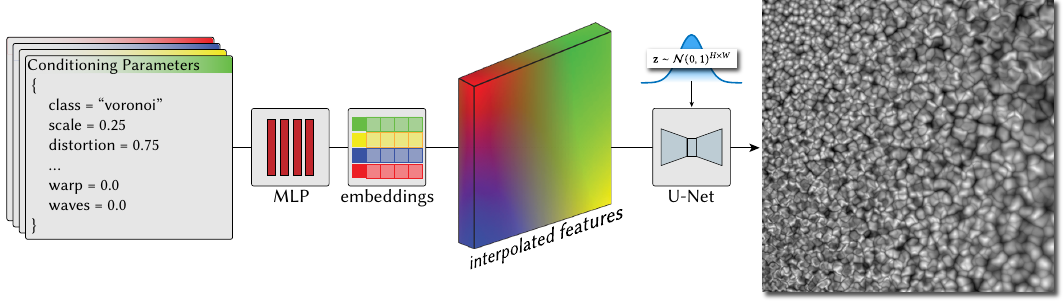}
    \caption{At inference time, we query our network using artificially constructed feature grids, enabling a flexible way to synthesize spatially-varying noise patterns. Here we embed four sets of noise parameters, pictorially shown as one of four colors. We blend the feature vectors using bilinear interpolation, creating a smoothly-varying feature grid, which our U-Net is able to transform into a Voronoi noise pattern with non-uniform scale and distortion characteristics.}
    \label{fig:inference}
\end{figure*}

\section{Method}\label{sec:method}
Given a collection of procedural noise textures sampled from various parametric noise generating functions, we seek to learn a conditional generative model that captures the noise textures given their accompanying parameter configurations, enabling the synthesis of a wide range of noises from a single universal function. Our conditional generative model will be formulated in a way that facilitates synthesizing spatially-varying noise textures (i.e. spatial blends between the categorical noise type as well as noise parameters), even though the training collection does not contain such samples. That is, we assume our given collection of noise samples only exhibits globally uniform semantic properties, as shown in Figure \ref{fig:dataset}.

We make use of a denoising diffusion probabilistic model (DDPM) that incorporates a spatially-varying conditioning mechanism. The model is trained using a data augmentation scheme that regulates the behavior of this conditioning module, ensuring that it behaves appropriately when given granular conditioning signals. In the following sections, we first detail the spatially-varying conditioning module, followed by the aforementioned data augmentation strategy. We defer details pertaining to our DDPM training to Section \ref{sec:implementation}.

\subsection{Spatially-varying conditioning}\label{sec:method_conditioning}
At training time, our conditioning signal is given by two vectors, $\*f_c$, a class embedding that encodes the categorical label of a particular noise type, and $\*f_p$, a list of parameters that determine semantic properties of the generated noise image (i.e. \texttt{scale}, \texttt{distortion}, etc.). Providing these parameters as explicit input to the model allows interpretable user control; it also disentangles control over the ``style'' of the generated noise (via these parameters) from the ``seed'' (i.e. stochastic component) of the noise functions via the initial Gaussian noise provided to the DDPM.

To facilitate synthesis of noise maps with spatially-varying properties, we employ spatially-adaptive denormalization (SPADE) \cite{park2019SPADE}. We first map the class and parameter vectors into a shared feature space with a small MLP, call the resulting feature vector $\*f$. The feature vector is tiled into a spatial grid, $\*F$, of shape $|\*f| \times H \times W$, where $H$, $W$ are the height and width respectively. This feature grid is used as input to the SPADE module. Following Dhariwal et al.~\shortcite{beatGANs}, we modulate the intermediate feature maps of our network's Group Normalization layers, making the SPADE conditioning function
\begin{displaymath}
\begin{gathered}
S(\*h, \*F) = \gamma(\*F) \odot \mathrm{GroupNorm}(\*h) + \beta(\*F) \\
\mathrm{GroupNorm}_g(\*h) = \frac{\*h_g - \mu_g}{\sqrt{\sigma_g^2 + \epsilon}}
\end{gathered}
\end{displaymath}
where $\gamma,\beta$ are convolutional layers that transform the feature grid into element-wise scales and shifts respectively, and $\*h$ is the incoming activations of the previous layer. The scales and shifts act on the output of a Group Normalization~\cite{group_norm} block, which is computed using the mean, $\mu$, and standard deviation, $\sigma$, of each group of channels $g$.

Our training data only contains noise textures that exhibit spatially \emph{uniform} properties, meaning that all feature entries in the grid $\*F$ are identical at training time. However, at inference time we may query the network with an artificially constructed feature grid -- for instance, we can spatially blend between multiple feature vectors to produce a conditioning signal that smoothly interpolates between noise types and noise parameters, as shown in Figure \ref{fig:inference}.
\subsubsection[Spherical class embeddings]{Spherical class embeddings\,\footnote{\label{publicationNote}The following subsection introduces an enhancement to our original method, identified post-publication, which improves results in many cases.}}\label{sec:sphericalEmbedding}
As previously mentioned, the network is conditioned on a class embedding, $\*f_c$, that is learned for each noise type during training. We find that regularizing this embedding space leads to substantial improvements when interpolating between classes. In particular, we penalize the deviation of the norms of these embeddings from a target norm. For embedding vectors of dimension $d$ that are initialized from $\*f_c\sim\mathcal{N}^d(0,\*I_d)$, we define the target norm, $T$, as the expected squared norm of a d-dimensional Gaussian vector according to the identity,
\begin{displaymath}
    T_d^n \coloneq \mathbb{E}_{\,\*f_c\sim\mathcal{N}^d(0,\*I_d)}\left[||\*f_c||_2^n\right] = 2^{n/2} \frac{\Gamma((d+n)/2)}{\Gamma(d/2)},
\end{displaymath}
where $\Gamma$ denotes Euler's gamma function. During training we then penalize the embeddings according to 
\begin{equation}
    \mathcal{L}_{\mathrm{reg}} = \frac{1}{|C|}\sum_{c\in C}(\|\*f_c\|_2^2 - T_d^2)^2,
\end{equation}
this imposes a \emph{spherical structure} on the embeddings, which means that interpolating between them can be done via spherical linear interpolation \cite{SLERP}. We find that imposing this structure greatly improves texture blending between classes. This is likely because the MLP (which acts on these embeddings) can now learn a smoother mapping into the final conditioning signal. As mentioned in note \ref{publicationNote}, this regularization was identified post-publication; hence, our primary results do not make use of this change. We demonstrate its effectiveness in Figure \ref{fig:slerpNoise} specifically.

\subsection{Enhancing localized conditioning}\label{method:cutmix}
Ideally, locally modifying the conditioning feature grid, $\*F$, within a confined region should correspondingly alter the generated noise texture solely within that region. However, our empirical observations revealed that such localized adjustments in the conditioning signal often lead to global (or near-global) changes in the resultant noise textures. This phenomenon can be primarily attributed to the architectural design of the employed U-Net neural network -- in particular, its bottleneck shape in combination with having many convolutional layers causes the receptive field of each output pixel to be relatively wide. This enables the conditioning signal to spatially propagate across the majority of the canvas, which is undesirable for the quality and controllability of our synthesized noise textures.

To address this, we incorporate a modified version of CutMix data augmentation \cite{yun2019cutmix} into our training procedure.
CutMix, a technique that is used to enhance performance in image classification tasks, involves creating synthetic training examples by stitching together axis-aligned crops of different dataset samples. Notably, the application of CutMix has been predominantly in classification tasks, with its use in generative models being relatively unexplored and confined to augmenting the performance of GAN discriminators~\cite{schonfeldCutMixApplication, huang2021CutMixApplication}. In our context, we apply CutMix by combining noise textures and their corresponding conditioning feature grids, as illustrated in Figure \ref{fig:pipeline}. This approach not only enriches the diversity of our training dataset, but also imparts a crucial capability to our model: the ability to respond correctly to spatially localized conditioning signals.

We train our network with CutMix data augmentation applied with a probability of $0.5$, i.e. for half of the training samples we train without applying CutMix. When augmenting a training image, we sample one to four (uniformly at random) auxiliary noise textures from the dataset, which are then randomly cropped (\emph{cut}) into rectangular patches with a randomly sampled rotation $\theta\sim\mathcal{U}(0,2\pi)$. Note that only the crop mask is rotated, not the texture itself. The resulting \emph{mixed} sample is then a composition of a base noise texture, and one to four texture patches. It is important that all sampled patches belong to unique noise types, otherwise the resulting training image would be invalid, that is, the image would no longer belong to the distribution that we are trying to capture.

\begin{figure}
    \centering
    \includegraphics[width=\linewidth, trim=0 14cm 0 0]{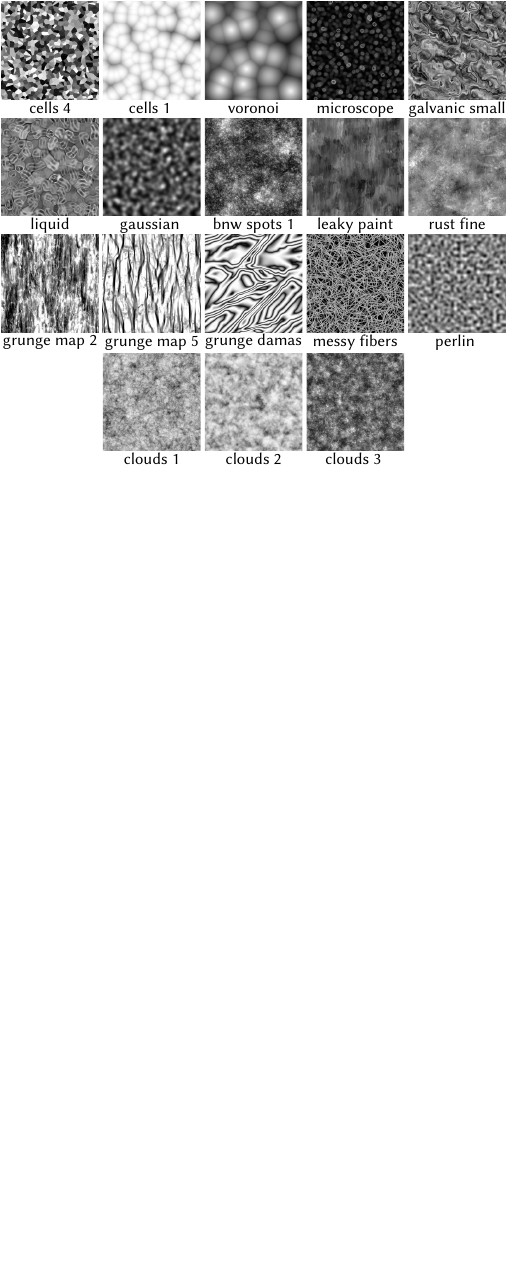}
    \caption{Samples from our noise dataset, procured from Adobe Substance 3D Designer. We sample \numNoiseTypes noises along with a variety of their parameters. Note that our dataset does not contain samples with spatially-varying properties.}
    \label{fig:dataset}
\end{figure}
\begin{figure*}[!ht]
    \centering
    \includegraphics[width=\linewidth]{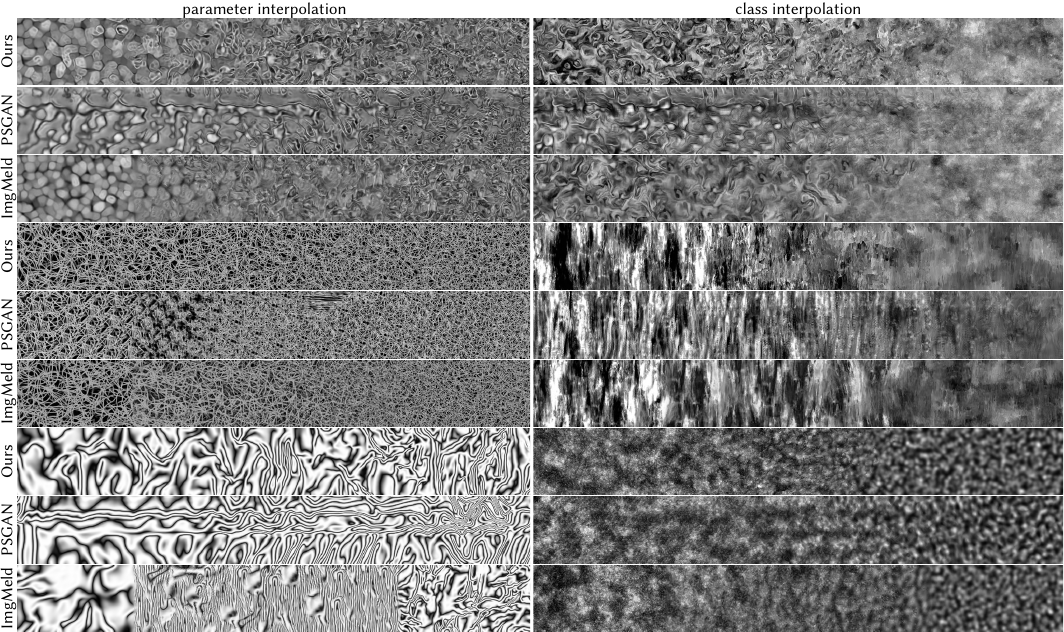}
    \caption{We compare our method to a neural texture synthesizer, PSGAN~\shortcite{PeriodicSpatialGAN}, as well as a non-parametric texture blending method, Image Melding~\shortcite{ImageMelding}. In the case of Image Melding, the first and last quarter of the image are given, only the remaining interior region is filled in. Both prior methods suffer from artifacts and repeated visual details, whereas our method is able to blend smoothly while synthesizing novel details throughout the canvas. We note that PSGAN produces anisotropic features that are not characteristic of the data distribution (e.g. horizontal streaks in bottom left example).}\label{fig:oursPSGANImageMeld}
\end{figure*}
\section{Implementation Details}\label{sec:implementation}
\paragraph{Noise dataset}
We procure a dataset of ${\sim}1.2$ million noise images, covering 18 unique noise functions, along with dense samplings of their respective parameter sets (listed in Appendix Table \ref{tab:noiseFunctions}). In particular, we sample the following noise functions from Adobe Substance 3D Designer: \texttt{cells 1, cells 4, voronoi, microscope view, grunge galvanic small, liquid, bnw spots 1, grunge leaky paint, grunge rust fine, grunge map 002, grunge map 005, grunge damas, messy fibers 3, perlin, gaussian, clouds 1, clouds 2, clouds 3}. A preview of our dataset is shown in Figure \ref{fig:dataset}. For each noise type, we deterministically sample $\num{16384}$ parameter sets using the low-discrepancy Halton number sequence, ensuring better coverage over the space of parameters. Each parameter set is sub-sampled four times, i.e. we query the noise functions with four different seeds for each parameter set, resulting in $\num{65536}$ image samples per noise type, and $\num{1179648}$ samples across the entire dataset. We sample at a resolution of $512\times512$px.

\paragraph{Network architecture}
Our U-Net model is composed of three spatial levels in the encoder and decoder, each level (including the bottleneck) consists of two ResNet blocks that are conditioned on the diffusion timestep, $t$, and the block's respective SPADE module. The two conditioning signals are composed together as
\begin{displaymath}
    \gamma_2(\*Z) \odot (\gamma_1(t) \cdot \mathrm{GroupNorm}(\*h) + \beta_1(t)) + \beta_2(\*Z)
\end{displaymath}
where $\gamma_1,\beta_1$ operate on $t$ and $\gamma_2,\beta_2$ belong to a SPADE block. Note that the former functions produce scalar quantities, whereas the latter functions produce spatial modulation maps. Time values are first encoded using sinusoidal positional encoding. The noise parameter MLP is defined by three layers of size 128. The parameter vector $\*f_p$ is a concatenation of all noise parameters present in the dataset, with entries zeroed out when not applicable. In total our U-Net has ${\sim}5.1$ million parameters.

\paragraph{Training}
Following Song et al.~\shortcite{song2020ddim} we formulate the U-Net as a noise predictor $\epsilon_\theta(\*x_t, t, \*f_c, \*f_p)$ making the training objective
\begin{equation}
    \mathcal{L} = \mathbb{E}_{\epsilon\sim\mathcal{N}(0,1),t\sim\mathcal{U}(0,1)} \| \epsilon - \epsilon_\theta(\*x_t, t, \*f_c, \*f_p) \|_2 + \lambda\mathcal{L}_{\mathrm{reg}}
    \label{eq:finalLoss}
\end{equation}
We make use of offset noise \cite{offsetNoise}, which replaces the noise sampling, $\epsilon\sim\mathcal{N}(0,1)$, with a slightly modified distribution, $\mathcal{N}(0.1\delta,1)$, where $\delta\sim\mathcal{N}(0,1)$. We find that offset noise greatly helps our network resolve noise types that are extremely dark or bright (see Appendix Figure \ref{fig:ablationOffsetNoise}). We also note that more principled alternatives to offset noise are now available~\cite{offsetNoiseAlternative}. In practice the timestep sampling, $t\sim\mathcal{U}(0,1)$, is discretized -- we make use of a standard number of timesteps, i.e. $\num{1000}$. We employ the cosine-beta schedule proposed by Nichol et al.~\shortcite{nichol2021improvedDiffusion}. The spherical embedding regularization loss is weighted by $\lambda=0.02$ in Equation \ref{eq:finalLoss}. Training is done in PyTorch; we use the AdamW optimizer \cite{loshchilov2017adamW} with a learning rate of $8\cdot10^{-5}$, $\beta_1=0.9, \beta_2=0.99$, and a weight decay of $0.01$. We train on 8 NVIDIA RTX 3090 GPUs for ${\sim}\num{300000}$ optimization steps with batches of size 128. Noise images are downsampled to $256\times256$px resolution for training.

\paragraph{Inference details \& performance.} Our model can perform 80 diffusion steps per second on a single NVIDIA RTX 3090 GPU at $256\times256$ resolution at full precision (fp32), scaling quadratically with the resolution. We use the DDIM sampler proposed by Song et al.~\shortcite{song2020ddim} -- unless otherwise specified, our noise figures and visualizations use 30 diffusion steps. To produce tileable noise maps, we modify all \texttt{conv2d} layers in our network to use a \emph{circular} padding mode, making the canvas topologically toroidal.

\section{Results and evaluation}\label{sec:results}
\begin{figure*}[p]
    \centering
    \includegraphics[width=\textwidth, trim=0 0.25cm 0 0]{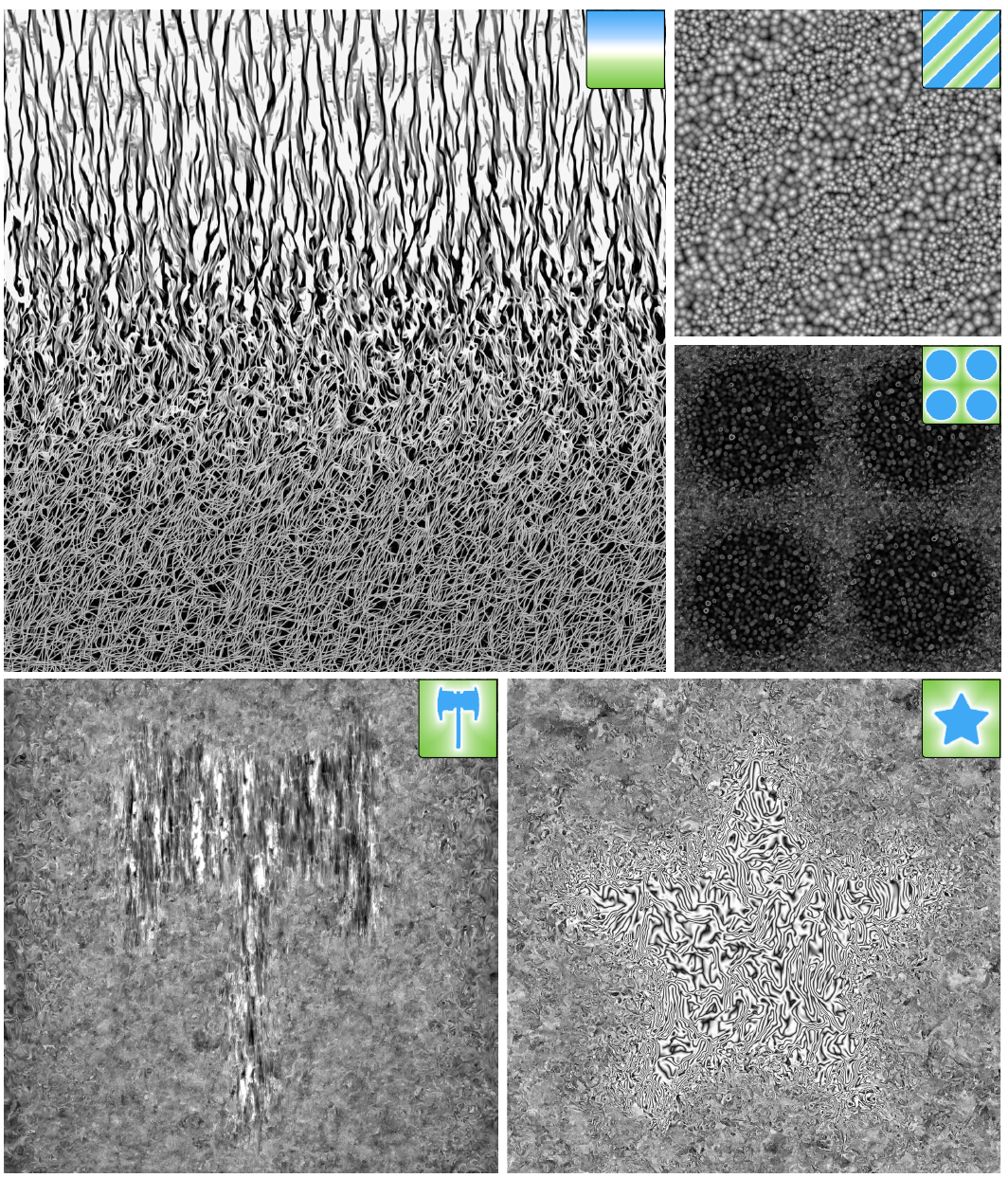}
    \caption{Several instances of spatially-varying noise, which arise from various blended feature grids. Blend maps are shown in the top right of each panel.}
    \label{fig:fullPage1_largeInterps}
\end{figure*}

\begin{figure*}[p]
    \centering
    \includegraphics[width=\textwidth]{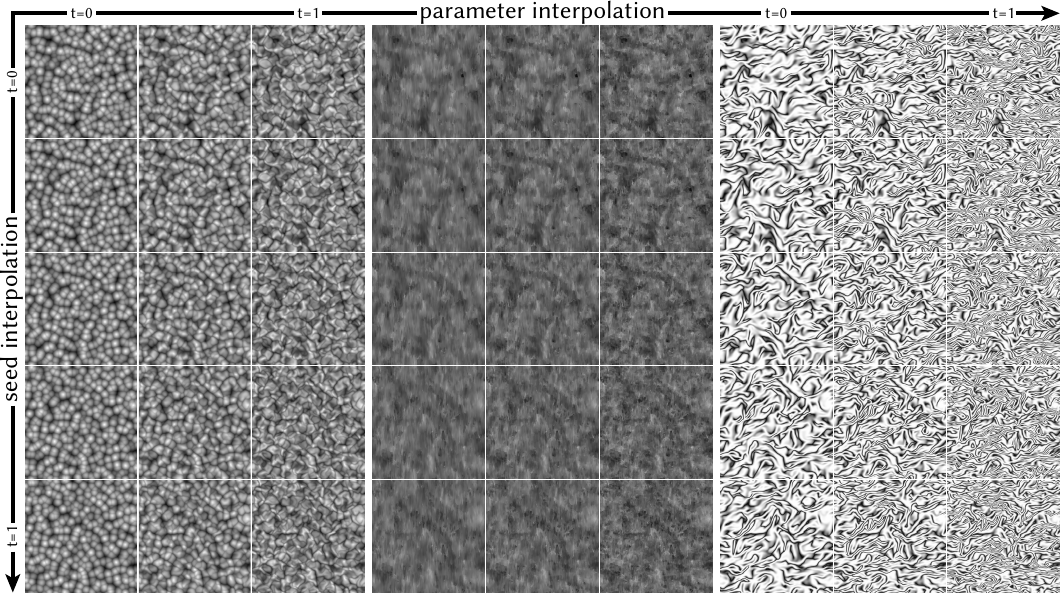}
    \vspace{-0.75cm}
    \caption{Our model is able to synthesize continuously-varying textures via interpolation of the conditioning parameters and diffusion noise (analogous to the ``seed'' in a traditional noise function). Here we show three noise types with 1) parameters interpolating (horizontal axis), and 2) the noise seed interpolating (vertical axis). We also note that, when using the same noise seed across different noise types, we observe that similar structures appear in the output -- we suggest zooming out to see this effect clearly. Noise interpolation videos are included in supplemental material.}
    \label{fig:fullPage2_interpolations}
\end{figure*}

\paragraph{Spatially and temporally varying noise}
We demonstrate the capabilities of our noise generator by synthesizing noise patterns using an assortment of blending maps, shown in Figures \ref{fig:teaser} and \ref{fig:fullPage1_largeInterps}. Our noise generator is able to be conditioned flexibly, allowing one to generate large noise patterns with diverse patterns and blends between them. We additionally show that our model is able to synthesize continuously-varying noise by interpolating the conditioning maps and/or diffusion noise (see Figure \ref{fig:fullPage2_interpolations} and supp. video). Finally, in Figure \ref{fig:oursPSGANImageMeld} we qualitatively compare our spatially-varying noise to a GAN-based texture synthesis method, PSGAN~\shortcite{PeriodicSpatialGAN}, as well as a non-parametric texture blending method, ImageMelding~\shortcite{ImageMelding}.
\begin{figure*}
    \centering
    \includegraphics[width=\linewidth]{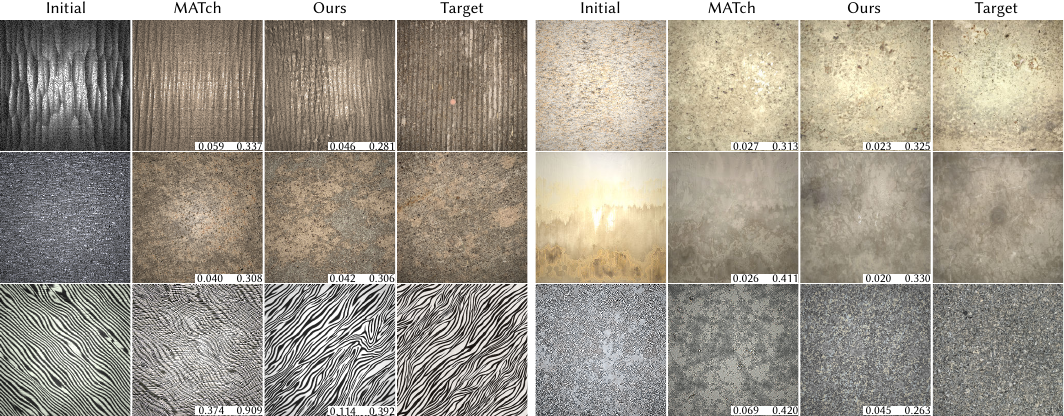}
    \caption{We demonstrate the utility of our method in material graph optimization. Given a procedural material graph and a target photograph, our noise generative model serves as a useful prior over the space of noise functions, facilitating the recovery of non-trivial patterns that are present in the target photos, and improving the baseline result given by MATch~\shortcite{Shi2020MATch}. Two similarity scores are reported for each result with respect to the target image: MATch's feature-based texture similarity metric (left) and LPIPS~\shortcite{LPIPS} (right). Lower is better for both scores.}
    \label{fig:diffMatResults}
\end{figure*}

\paragraph{Quantitative evaluation} We evaluate our noise generator's spatially-uniform noise textures against the ground truth textures from Adobe Substance 3D Designer using Frechet Inception Distance (FID), which measures the distributional similarity between synthetic samples and the data distribution \cite{FID}. We sample \num{20000} synthetic images for each noise type and evaluate the FID of each noise type separately. FID scores are reported in Appendix Table \ref{tab:fidMain}, alongside analogous scores for PSGAN. Our method achieves a mean FID of 20.9 and median of 13.1, while PSGAN scores 99.2 and 87.5 respectively (lower is better). We note that, for many noise types, PSGAN suffers from mode collapse, which causes virtually all of the synthesized results to have similar visual features and hence its FID is severely impacted. As mentioned in Section \ref{sec:relatedWorkTextureSynth}, GAN-based methods struggle to faithfully capture our dataset due to the presence of many disjoint data modes~\cite{xiao2022tackling}. Finally, we show additional results with modified U-Net architectures in Appendix \ref{sec:supp_Architectures}.

\begin{figure}
    \centering
    \includegraphics[width=\linewidth]{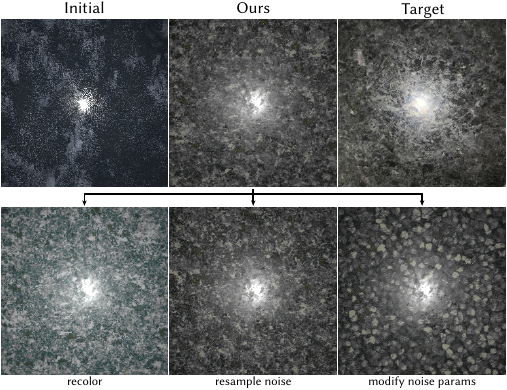}
    \caption{After optimization, we can easily modify the resulting procedural material graphs. Here a marble material graph is optimized (first row), and edited in various ways (second row). We show the results of editing the noise colorization, resampling our diffusion noise, and modifying our model's conditioning inputs.}
    \label{fig:graphEdit}
\end{figure}

\paragraph{Inverse material graph design}
We incorporate our noise generator into the differentiable material graph library MATch~\cite{Shi2020MATch}. Given a target photo, we select a template material graph from a selection of 88 graphs using the MATch-provided texture descriptors. In the graph, we replace one noise generator node with our model, and expose optimizable parameters for the class vector $\*f_c$, parameter vector $\*f_p$, and the diffusion model's latent noise $\*z$. We include an L1 regularization on the former vectors to encourage sparsity in the optimized parameters. In Figure \ref{fig:diffMatResults} we show that our noise generator provides a valuable prior over the space of noise functions, allowing end-to-end optimization to recover more accurate reconstructions of the target photo. We additionally show results of editing operations on an optimized material graph in Figure \ref{fig:graphEdit}. Target photos are sampled from the Glossy dataset by Zhou et al.~\shortcite{zhou2023photomat}. Further details for this application are in Appendix \ref{supp:InverseDesign}.

\begin{figure*}[p]
    \centering
    \includegraphics[width=\textwidth]{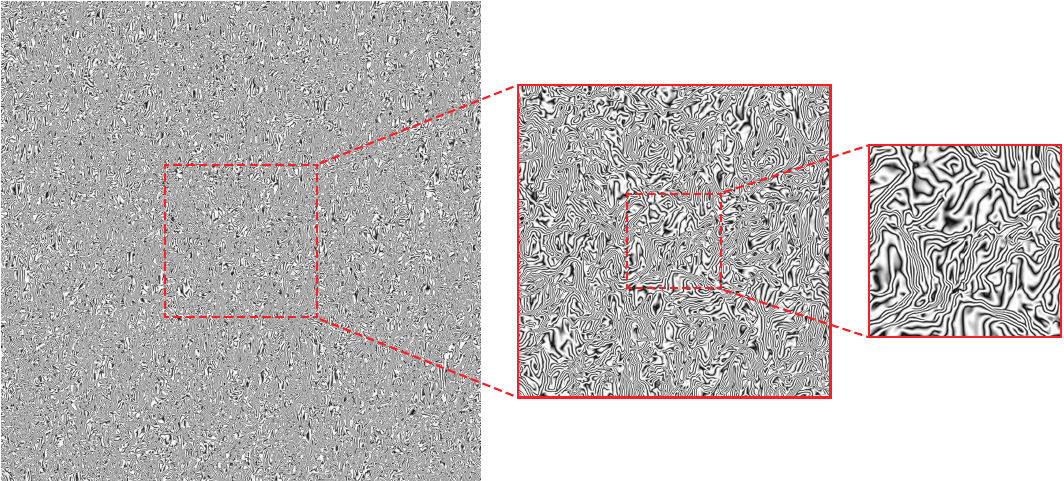}
    \caption{Our model is capable of synthesizing noise maps at resolutions that far exceed the training resolution, thus facilitating one to paint ``infinite'' canvases. Here we see a $2048\times2048$px Damascus steel noise pattern with intricate fine details, which was generated via a single diffusion process. Please zoom into the figure for full details.}
    \label{fig:fullPage2_infiniteImage}
\end{figure*}

\begin{figure}
    \centering
    \includegraphics[width=\linewidth]{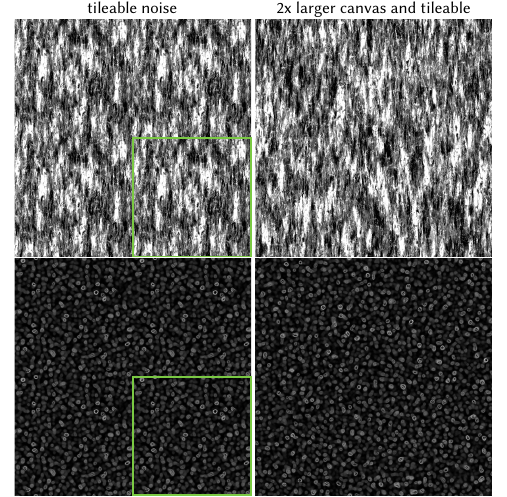}
    \caption{Our model is able to produce seamless tileable images, as shown by the green regions, which are tiled into 2-by-2 noise maps. We also support synthesizing noise on an arbitrary canvas, avoiding noticeable repeated patterns, which is a common problem with tileable textures.}
    \label{fig:tileInfinite}
\end{figure}

\paragraph{Tileability and texture size agnosticism}
Our model is able to produce noise images that are tileable, which is crucial for many downstream graphics applications. Additionally, we are able to synthesize noise at arbitrary sizes (within computational limits). In tandem, these two properties are highly desired, as it allows one to generate large tileable images that do not contain excessive repeated visual patterns, which is a common problem with tileable images. We  demonstrate these capabilities in Figure \ref{fig:tileInfinite}, where we show tileable \texttt{grunge map 5} and \texttt{microscope} noise images, as well as the same noises synthesized on canvases that are twice as large, eliminating the repeated visual content. Additionally, in Figure \ref{fig:fullPage2_infiniteImage} we show a much larger \texttt{Damascus} pattern that fills a $2048\times2048$ canvas -- we highlight that fine-grained details are not at this size.

\begin{figure}
    \centering
    \includegraphics[width=\linewidth]{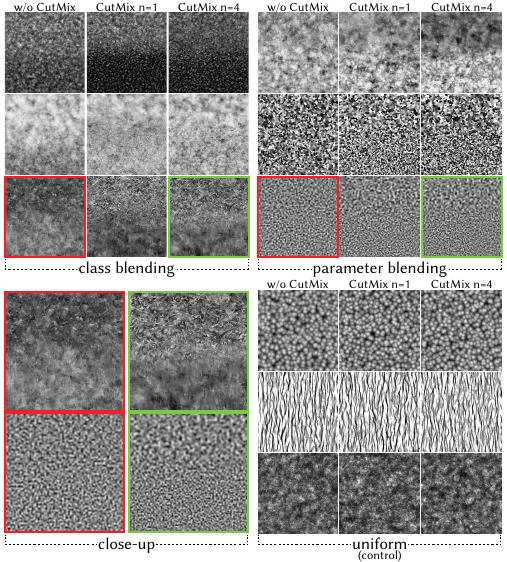}
    \caption{Ablation study of CutMix data augmentation. Without CutMix, the U-Net fails to resolve noise maps that contain non-uniform characteristics. For instance, in row 3 of the \emph{class blending} panel, the network cannot adequately blend between noise classes, causing the \texttt{galvanic} noise pattern to disappear entirely. We include models with one and four applications of CutMix, as detailed in Section \ref{sec:results}. Finally, the \emph{uniform} panel (no blending) acts as a control group -- as expected, all outputs are similar.}
    \label{fig:ablationCutMix}
\end{figure}

\paragraph{CutMix augmentation}
We demonstrate the affect of our modified CutMix data augmentation strategy by training models with: no augmentation, augmentation with only one patch (\texttt{CutMix $n=1$}), and augmentation with one to four patches sampled uniformly at random (\emph{CutMix $n\in[1,4]$}). In Figure \ref{fig:ablationCutMix}, we see that removing the augmentation strategy severely hinders our model's ability to respond to the conditioning signal in a local manner, producing noise patterns whose characteristics are not well-pronounced. Intuitively, this is because the conditioning signal becomes ``blurred'' as it propagates through the network's layers. Due to the lack of meaningful ground-truth data, we qualitatively observe that \texttt{Cutmix $n=1$} and \texttt{Cutmix $n=4$} produce similar results, with the latter being able to interpolate noise characteristics more smoothly in some cases.

\section{Conclusion}

\begin{figure}
    \centering
    \includegraphics[width=\linewidth]{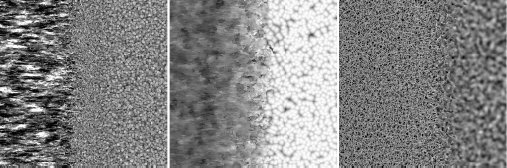}
    \caption{Our method is unable to smoothly blend between some noise types. In the configurations above, we see artifacts in the transition regions.}
    \label{fig:limitations}
    \vspace{-17pt}
\end{figure}

We presented a new method for learning a single generative model for a wide range of procedural noise types.
The continuous space of noise patterns learned by our model allows for smooth blending between noise types; the characteristics of the output noise are still controllable by interpretable parameters provided as input to the model.
In addition, our novel approach for spatially-varying conditioning allows the model to generate spatially-varying blends between different noises despite never having seen such data at training time.
We used this model to produce a variety of visually compelling textures, and we showed a proof-of-concept application of how it can be used to improve inverse procedural material design.

Our model is not without limitations.
Not all pairs of noise types can be interpolated well: if the geometric features of two noise patterns are too dissimilar, the intermediate regions of a blend between them can look blurry or otherwise visually awkward (see Fig.~\ref{fig:limitations}).
It may be possible to reduce these artefacts with further improvements to our CutMix-based data augmentation (e.g. non-rectangular patches; adaptive patch sampling to focus more training time on ``difficult'' transitions).
But to some extent, such behavior is inevitable, and finding good noise pairs for interpolation is an artistic decision.
It may be possible to leverage the structure of our noise embedding space to suggest good candidates for blending. Finally, diffusion models are known to model low-density modes of the data distribution less accurately~\cite{yangSongDiffusionBlog,diffusionLowDensity}; in Appendix \ref{sec:supp_additionalNoise} we demonstrate our model on additional noise types, and observe that some low-density modes of these noise functions are poorly captured. More careful data sampling techniques may be needed when the desired noise distribution exhibits such modes; e.g. by sampling some regions of the parameter space more frequently.

Procedural material authoring tools such as Adobe Substance Designer often have deterministic pattern generators in addition to stochastic noise generators.
Could our model include these in its learned space of blendable output?
The deterministic nature of these patterns makes them an unnatural fit for generation via denoising diffusion model; modifications to our model and/or training procedure would be necessary to achieve this goal.

It would also be interesting to explore more techniques for interactively creating spatially-vary noise textures.
For example, one might consider an interaction metaphor where the user places `droplets' of noise on a canvas, and those droplets spread an interact with one another by e.g. solving a diffusion equation.

Finally, the application to inverse procedural material design that we presented is an early proof-of-concept; much more could be done in this direction.
Distilling our model into a one-shot diffusion model~\cite{liu2023instaflow} would make it tractable to replace \emph{every} noise generator node in a procedural material graph with our model.
Provided other graph operations are differentiable, this would open up the possibility of solving for the structure of a material graph via continuous optimization: initializing the graph to be over-complete, assigning weights to edges, and imposing a sparsity prior on edge weights.
Similar approaches have been successfully employed for inferring procedural representations of 3D shapes~\cite{kania2020ucsgnet,ren2021csg}.
In addition, allowing the optimization of spatially-varying noise generators within material graphs might allow for recovering smaller, easier-to-use graphs while still retaining the benefits of interpretable parameters.
\begin{figure}
    \centering
    \includegraphics[width=\linewidth]{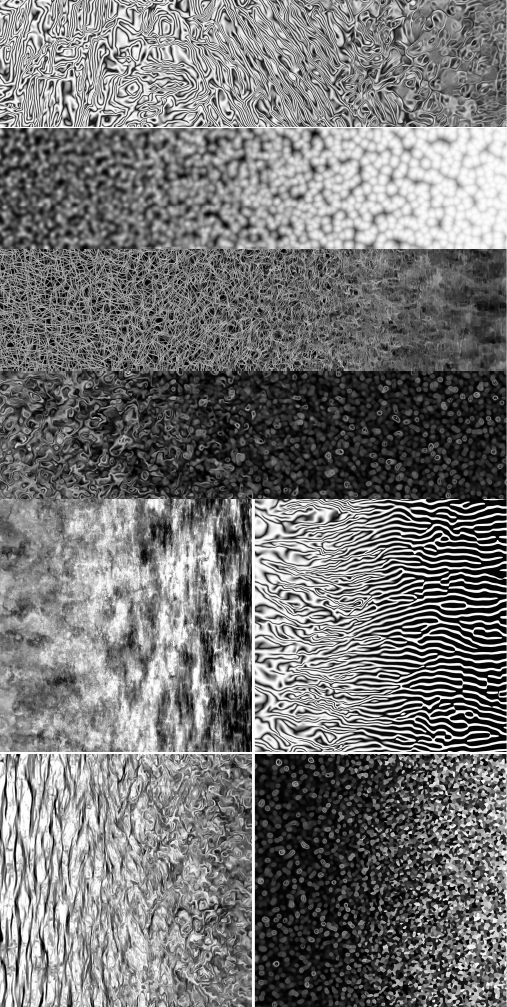}
    \caption{Noise interpolations that arise from our spherically regularized embedding space, as described in Section \ref{sec:sphericalEmbedding}. The blending between noise types appears organic and is largely free of the artifacts noted in Figure \ref{fig:limitations}.}
    \label{fig:slerpNoise}
\end{figure}

\begin{acks}
This material is based upon work that was supported by the National Science Foundation Graduate Research Fellowship under Grant No. 2040433. Part of this work was done while Arman Maesumi was an intern at Adobe Research.
\end{acks}

\bibliographystyle{ACM-Reference-Format}
\bibliography{mybib}


\begin{thebibliography}{58}


\ifx \showCODEN    \undefined \def \showCODEN     #1{\unskip}     \fi
\ifx \showDOI      \undefined \def \showDOI       #1{#1}\fi
\ifx \showISBNx    \undefined \def \showISBNx     #1{\unskip}     \fi
\ifx \showISBNxiii \undefined \def \showISBNxiii  #1{\unskip}     \fi
\ifx \showISSN     \undefined \def \showISSN      #1{\unskip}     \fi
\ifx \showLCCN     \undefined \def \showLCCN      #1{\unskip}     \fi
\ifx \shownote     \undefined \def \shownote      #1{#1}          \fi
\ifx \showarticletitle \undefined \def \showarticletitle #1{#1}   \fi
\ifx \showURL      \undefined \def \showURL       {\relax}        \fi
\providecommand\bibfield[2]{#2}
\providecommand\bibinfo[2]{#2}
\providecommand\natexlab[1]{#1}
\providecommand\showeprint[2][]{arXiv:#2}

\bibitem[Adobe(2023a)]%
        {adobeSubstance}
\bibfield{author}{\bibinfo{person}{Adobe}.} \bibinfo{year}{2023}\natexlab{a}.
\newblock \bibinfo{title}{Adobe Substance 3D Designer}.
\newblock \bibinfo{howpublished}{\url{https://www.adobe.com/products/substance3d-designer.html}}.
\newblock


\bibitem[Adobe(2023b)]%
        {substanceDocumentation}
\bibfield{author}{\bibinfo{person}{Adobe}.} \bibinfo{year}{2023}\natexlab{b}.
\newblock \bibinfo{title}{Adobe Substance 3D Documentation}.
\newblock \bibinfo{howpublished}{\url{https://helpx.adobe.com/substance-3d-designer/substance-compositing-graphs/nodes-reference-for-substance-compositing-graphs/node-library.html}}.
\newblock


\bibitem[Arjovsky et~al\mbox{.}(2017)]%
        {WGAN}
\bibfield{author}{\bibinfo{person}{Martin Arjovsky}, \bibinfo{person}{Soumith Chintala}, {and} \bibinfo{person}{L{\'e}on Bottou}.} \bibinfo{year}{2017}\natexlab{}.
\newblock \showarticletitle{{W}asserstein Generative Adversarial Networks}. In \bibinfo{booktitle}{\emph{Proceedings of the 34th International Conference on Machine Learning}} \emph{(\bibinfo{series}{Proceedings of Machine Learning Research}, Vol.~\bibinfo{volume}{70})}, \bibfield{editor}{\bibinfo{person}{Doina Precup} {and} \bibinfo{person}{Yee~Whye Teh}} (Eds.). \bibinfo{publisher}{PMLR}, \bibinfo{pages}{214--223}.
\newblock
\urldef\tempurl%
\url{https://proceedings.mlr.press/v70/arjovsky17a.html}
\showURL{%
\tempurl}


\bibitem[Bergmann et~al\mbox{.}(2017)]%
        {PeriodicSpatialGAN}
\bibfield{author}{\bibinfo{person}{Urs Bergmann}, \bibinfo{person}{Nikolay Jetchev}, {and} \bibinfo{person}{Roland Vollgraf}.} \bibinfo{year}{2017}\natexlab{}.
\newblock \showarticletitle{Learning Texture Manifolds with the Periodic Spatial {GAN}}. In \bibinfo{booktitle}{\emph{Proceedings of the 34th International Conference on Machine Learning}} \emph{(\bibinfo{series}{Proceedings of Machine Learning Research}, Vol.~\bibinfo{volume}{70})}, \bibfield{editor}{\bibinfo{person}{Doina Precup} {and} \bibinfo{person}{Yee~Whye Teh}} (Eds.). \bibinfo{publisher}{PMLR}, \bibinfo{pages}{469--477}.
\newblock
\urldef\tempurl%
\url{https://proceedings.mlr.press/v70/bergmann17a.html}
\showURL{%
\tempurl}


\bibitem[Blender(2023)]%
        {blender}
\bibfield{author}{\bibinfo{person}{Blender}.} \bibinfo{year}{2023}\natexlab{}.
\newblock \bibinfo{title}{Blender}.
\newblock \bibinfo{howpublished}{\url{https://www.blender.org}}.
\newblock


\bibitem[Cook and DeRose(2005)]%
        {cook2005wavelet}
\bibfield{author}{\bibinfo{person}{Robert~L Cook} {and} \bibinfo{person}{Tony DeRose}.} \bibinfo{year}{2005}\natexlab{}.
\newblock \showarticletitle{Wavelet noise}.
\newblock \bibinfo{journal}{\emph{ACM Transactions on Graphics (TOG)}} \bibinfo{volume}{24}, \bibinfo{number}{3} (\bibinfo{year}{2005}), \bibinfo{pages}{803--811}.
\newblock


\bibitem[Darabi et~al\mbox{.}(2012)]%
        {ImageMelding}
\bibfield{author}{\bibinfo{person}{Soheil Darabi}, \bibinfo{person}{Eli Shechtman}, \bibinfo{person}{Connelly Barnes}, \bibinfo{person}{Dan~B Goldman}, {and} \bibinfo{person}{Pradeep Sen}.} \bibinfo{year}{2012}\natexlab{}.
\newblock \showarticletitle{{I}mage {M}elding: Combining Inconsistent Images using Patch-based Synthesis}.
\newblock \bibinfo{journal}{\emph{ACM Transactions on Graphics (TOG) (Proceedings of SIGGRAPH 2012)}} \bibinfo{volume}{31}, \bibinfo{number}{4}, Article \bibinfo{articleno}{82} (\bibinfo{year}{2012}), \bibinfo{numpages}{82:1--82:10}~pages.
\newblock


\bibitem[Dhariwal and Nichol(2021)]%
        {beatGANs}
\bibfield{author}{\bibinfo{person}{Prafulla Dhariwal} {and} \bibinfo{person}{Alexander Nichol}.} \bibinfo{year}{2021}\natexlab{}.
\newblock \showarticletitle{Diffusion models beat gans on image synthesis}.
\newblock \bibinfo{journal}{\emph{Advances in neural information processing systems}}  \bibinfo{volume}{34} (\bibinfo{year}{2021}), \bibinfo{pages}{8780--8794}.
\newblock


\bibitem[Dorsey and Hanrahany(2006)]%
        {dorsey2006Patinas}
\bibfield{author}{\bibinfo{person}{Julie Dorsey} {and} \bibinfo{person}{Pat Hanrahany}.} \bibinfo{year}{2006}\natexlab{}.
\newblock \showarticletitle{Modeling and rendering of metallic patinas}.
\newblock In \bibinfo{booktitle}{\emph{ACM SIGGRAPH 2006 Courses}}. \bibinfo{pages}{2--es}.
\newblock


\bibitem[Ebert et~al\mbox{.}(2002)]%
        {musgrave2002texturingBook}
\bibfield{author}{\bibinfo{person}{David Ebert}, \bibinfo{person}{Kenton Musgrave}, \bibinfo{person}{Darwyn Peachey}, \bibinfo{person}{Ken Perlin}, {and} \bibinfo{person}{Steve Worley}.} \bibinfo{year}{2002}\natexlab{}.
\newblock \bibinfo{booktitle}{\emph{Texturing And Modeling. A Procedural Approach} (\bibinfo{edition}{3} ed.)}.
\newblock \bibinfo{publisher}{Morgan Kaufmann}.
\newblock


\bibitem[Fournier et~al\mbox{.}(1982)]%
        {fournier1982subdivision}
\bibfield{author}{\bibinfo{person}{Alain Fournier}, \bibinfo{person}{Don Fussell}, {and} \bibinfo{person}{Loren Carpenter}.} \bibinfo{year}{1982}\natexlab{}.
\newblock \showarticletitle{Computer rendering of stochastic models}.
\newblock \bibinfo{journal}{\emph{Commun. ACM}} \bibinfo{volume}{25}, \bibinfo{number}{6} (\bibinfo{year}{1982}), \bibinfo{pages}{371--384}.
\newblock


\bibitem[Galin et~al\mbox{.}(2019)]%
        {galin2019terrainModelingReview}
\bibfield{author}{\bibinfo{person}{Eric Galin}, \bibinfo{person}{Eric Gu{\'e}rin}, \bibinfo{person}{Adrien Peytavie}, \bibinfo{person}{Guillaume Cordonnier}, \bibinfo{person}{Marie-Paule Cani}, \bibinfo{person}{Bedrich Benes}, {and} \bibinfo{person}{James Gain}.} \bibinfo{year}{2019}\natexlab{}.
\newblock \showarticletitle{A review of digital terrain modeling}. In \bibinfo{booktitle}{\emph{Computer Graphics Forum}}, Vol.~\bibinfo{volume}{38}. Wiley Online Library, \bibinfo{pages}{553--577}.
\newblock


\bibitem[Gatys et~al\mbox{.}(2015)]%
        {gatys2015texture}
\bibfield{author}{\bibinfo{person}{Leon~A. Gatys}, \bibinfo{person}{Alexander~S. Ecker}, {and} \bibinfo{person}{Matthias Bethge}.} \bibinfo{year}{2015}\natexlab{}.
\newblock \showarticletitle{Texture Synthesis Using Convolutional Neural Networks}. In \bibinfo{booktitle}{\emph{NeurIPS}}.
\newblock


\bibitem[G{\'e}nevaux et~al\mbox{.}(2013)]%
        {genevaux2013terrain}
\bibfield{author}{\bibinfo{person}{Jean-David G{\'e}nevaux}, \bibinfo{person}{{\'E}ric Galin}, \bibinfo{person}{Eric Gu{\'e}rin}, \bibinfo{person}{Adrien Peytavie}, {and} \bibinfo{person}{Bedrich Benes}.} \bibinfo{year}{2013}\natexlab{}.
\newblock \showarticletitle{Terrain generation using procedural models based on hydrology}.
\newblock \bibinfo{journal}{\emph{ACM Transactions on Graphics (TOG)}} \bibinfo{volume}{32}, \bibinfo{number}{4} (\bibinfo{year}{2013}), \bibinfo{pages}{1--13}.
\newblock


\bibitem[Guerrero et~al\mbox{.}(2022)]%
        {guerrero2022matformer}
\bibfield{author}{\bibinfo{person}{Paul Guerrero}, \bibinfo{person}{Milo{\v{s}} Ha{\v{s}}an}, \bibinfo{person}{Kalyan Sunkavalli}, \bibinfo{person}{Radom{\'\i}r M{\v{e}}ch}, \bibinfo{person}{Tamy Boubekeur}, {and} \bibinfo{person}{Niloy~J Mitra}.} \bibinfo{year}{2022}\natexlab{}.
\newblock \showarticletitle{MatFormer: A generative model for procedural materials}.
\newblock \bibinfo{journal}{\emph{arXiv preprint arXiv:2207.01044}} (\bibinfo{year}{2022}).
\newblock


\bibitem[Guo et~al\mbox{.}(2020)]%
        {guo2020bayesianProcMat}
\bibfield{author}{\bibinfo{person}{Yu Guo}, \bibinfo{person}{Milo{\v{s}} Ha{\v{s}}an}, \bibinfo{person}{Lingqi Yan}, {and} \bibinfo{person}{Shuang Zhao}.} \bibinfo{year}{2020}\natexlab{}.
\newblock \showarticletitle{A bayesian inference framework for procedural material parameter estimation}. In \bibinfo{booktitle}{\emph{Computer Graphics Forum}}, Vol.~\bibinfo{volume}{39}. Wiley Online Library, \bibinfo{pages}{255--266}.
\newblock


\bibitem[Guttenberg(2023)]%
        {offsetNoise}
\bibfield{author}{\bibinfo{person}{Nicholas Guttenberg}.} \bibinfo{year}{2023}\natexlab{}.
\newblock \showarticletitle{Diffusion with Offset Noise}.
\newblock  (\bibinfo{year}{2023}).
\newblock


\bibitem[Heusel et~al\mbox{.}(2017)]%
        {FID}
\bibfield{author}{\bibinfo{person}{Martin Heusel}, \bibinfo{person}{Hubert Ramsauer}, \bibinfo{person}{Thomas Unterthiner}, \bibinfo{person}{Bernhard Nessler}, {and} \bibinfo{person}{Sepp Hochreiter}.} \bibinfo{year}{2017}\natexlab{}.
\newblock \showarticletitle{Gans trained by a two time-scale update rule converge to a local nash equilibrium}.
\newblock \bibinfo{journal}{\emph{Advances in neural information processing systems}}  \bibinfo{volume}{30} (\bibinfo{year}{2017}).
\newblock


\bibitem[Hinsinger et~al\mbox{.}(2002)]%
        {hinsinger2002interactiveOceans}
\bibfield{author}{\bibinfo{person}{Damien Hinsinger}, \bibinfo{person}{Fabrice Neyret}, {and} \bibinfo{person}{Marie-Paule Cani}.} \bibinfo{year}{2002}\natexlab{}.
\newblock \showarticletitle{Interactive animation of ocean waves}. In \bibinfo{booktitle}{\emph{Proceedings of the 2002 ACM SIGGRAPH/Eurographics symposium on Computer animation}}. \bibinfo{pages}{161--166}.
\newblock


\bibitem[Ho et~al\mbox{.}(2020)]%
        {ho2020ddpm}
\bibfield{author}{\bibinfo{person}{Jonathan Ho}, \bibinfo{person}{Ajay Jain}, {and} \bibinfo{person}{Pieter Abbeel}.} \bibinfo{year}{2020}\natexlab{}.
\newblock \showarticletitle{Denoising diffusion probabilistic models}.
\newblock \bibinfo{journal}{\emph{Advances in neural information processing systems}}  \bibinfo{volume}{33} (\bibinfo{year}{2020}), \bibinfo{pages}{6840--6851}.
\newblock


\bibitem[Hu et~al\mbox{.}(2019)]%
        {hu2019InverseProc}
\bibfield{author}{\bibinfo{person}{Yiwei Hu}, \bibinfo{person}{Julie Dorsey}, {and} \bibinfo{person}{Holly Rushmeier}.} \bibinfo{year}{2019}\natexlab{}.
\newblock \showarticletitle{A novel framework for inverse procedural texture modeling}.
\newblock \bibinfo{journal}{\emph{ACM Transactions on Graphics (ToG)}} \bibinfo{volume}{38}, \bibinfo{number}{6} (\bibinfo{year}{2019}), \bibinfo{pages}{1--14}.
\newblock


\bibitem[Hu et~al\mbox{.}(2022)]%
        {hu2022diffProxy}
\bibfield{author}{\bibinfo{person}{Yiwei Hu}, \bibinfo{person}{Paul Guerrero}, \bibinfo{person}{Milos Hasan}, \bibinfo{person}{Holly Rushmeier}, {and} \bibinfo{person}{Valentin Deschaintre}.} \bibinfo{year}{2022}\natexlab{}.
\newblock \showarticletitle{Node graph optimization using differentiable proxies}. In \bibinfo{booktitle}{\emph{ACM SIGGRAPH 2022 conference proceedings}}. \bibinfo{pages}{1--9}.
\newblock


\bibitem[Huang et~al\mbox{.}(2021)]%
        {huang2021CutMixApplication}
\bibfield{author}{\bibinfo{person}{Zhizhong Huang}, \bibinfo{person}{Junping Zhang}, \bibinfo{person}{Yi Zhang}, {and} \bibinfo{person}{Hongming Shan}.} \bibinfo{year}{2021}\natexlab{}.
\newblock \showarticletitle{DU-GAN: Generative adversarial networks with dual-domain U-Net-based discriminators for low-dose CT denoising}.
\newblock \bibinfo{journal}{\emph{IEEE Transactions on Instrumentation and Measurement}}  \bibinfo{volume}{71} (\bibinfo{year}{2021}), \bibinfo{pages}{1--12}.
\newblock


\bibitem[Jones and Poggio(1998)]%
        {morphableModels}
\bibfield{author}{\bibinfo{person}{Michael~J Jones} {and} \bibinfo{person}{Tomaso Poggio}.} \bibinfo{year}{1998}\natexlab{}.
\newblock \showarticletitle{Multidimensional morphable models}. In \bibinfo{booktitle}{\emph{Sixth International Conference on Computer Vision (IEEE Cat. No. 98CH36271)}}. IEEE, \bibinfo{pages}{683--688}.
\newblock


\bibitem[Kania et~al\mbox{.}(2020)]%
        {kania2020ucsgnet}
\bibfield{author}{\bibinfo{person}{Kacper Kania}, \bibinfo{person}{Maciej Zieba}, {and} \bibinfo{person}{Tomasz Kajdanowicz}.} \bibinfo{year}{2020}\natexlab{}.
\newblock \showarticletitle{UCSG-NET- Unsupervised Discovering of Constructive Solid Geometry Tree}. In \bibinfo{booktitle}{\emph{Advances in Neural Information Processing Systems}}.
\newblock


\bibitem[Lagae et~al\mbox{.}(2010)]%
        {lagae2010noiseSurvey}
\bibfield{author}{\bibinfo{person}{Ares Lagae}, \bibinfo{person}{Sylvain Lefebvre}, \bibinfo{person}{Rob Cook}, \bibinfo{person}{Tony DeRose}, \bibinfo{person}{George Drettakis}, \bibinfo{person}{David~S Ebert}, \bibinfo{person}{John~P Lewis}, \bibinfo{person}{Ken Perlin}, {and} \bibinfo{person}{Matthias Zwicker}.} \bibinfo{year}{2010}\natexlab{}.
\newblock \showarticletitle{A survey of procedural noise functions}. In \bibinfo{booktitle}{\emph{Computer Graphics Forum}}, Vol.~\bibinfo{volume}{29}. Wiley Online Library, \bibinfo{pages}{2579--2600}.
\newblock


\bibitem[Lagae et~al\mbox{.}(2009)]%
        {lagae2009gabor}
\bibfield{author}{\bibinfo{person}{Ares Lagae}, \bibinfo{person}{Sylvain Lefebvre}, \bibinfo{person}{George Drettakis}, {and} \bibinfo{person}{Philip Dutr{\'e}}.} \bibinfo{year}{2009}\natexlab{}.
\newblock \showarticletitle{Procedural noise using sparse Gabor convolution}.
\newblock \bibinfo{journal}{\emph{ACM Transactions on Graphics (TOG)}} \bibinfo{volume}{28}, \bibinfo{number}{3} (\bibinfo{year}{2009}), \bibinfo{pages}{1--10}.
\newblock


\bibitem[Li and Wand(2016)]%
        {MarkovianGAN}
\bibfield{author}{\bibinfo{person}{Chuan Li} {and} \bibinfo{person}{Michael Wand}.} \bibinfo{year}{2016}\natexlab{}.
\newblock \showarticletitle{Precomputed Real-Time Texture Synthesis with Markovian Generative Adversarial Networks}. In \bibinfo{booktitle}{\emph{Computer Vision -- ECCV 2016}}, \bibfield{editor}{\bibinfo{person}{Bastian Leibe}, \bibinfo{person}{Jiri Matas}, \bibinfo{person}{Nicu Sebe}, {and} \bibinfo{person}{Max Welling}} (Eds.). \bibinfo{publisher}{Springer International Publishing}, \bibinfo{address}{Cham}, \bibinfo{pages}{702--716}.
\newblock
\showISBNx{978-3-319-46487-9}


\bibitem[Lin et~al\mbox{.}(2024)]%
        {offsetNoiseAlternative}
\bibfield{author}{\bibinfo{person}{Shanchuan Lin}, \bibinfo{person}{Bingchen Liu}, \bibinfo{person}{Jiashi Li}, {and} \bibinfo{person}{Xiao Yang}.} \bibinfo{year}{2024}\natexlab{}.
\newblock \showarticletitle{Common diffusion noise schedules and sample steps are flawed}. In \bibinfo{booktitle}{\emph{Proceedings of the IEEE/CVF Winter Conference on Applications of Computer Vision}}. \bibinfo{pages}{5404--5411}.
\newblock


\bibitem[Liu et~al\mbox{.}(2023)]%
        {liu2023instaflow}
\bibfield{author}{\bibinfo{person}{Xingchao Liu}, \bibinfo{person}{Xiwen Zhang}, \bibinfo{person}{Jianzhu Ma}, \bibinfo{person}{Jian Peng}, {and} \bibinfo{person}{Qiang Liu}.} \bibinfo{year}{2023}\natexlab{}.
\newblock \bibinfo{title}{InstaFlow: One Step is Enough for High-Quality Diffusion-Based Text-to-Image Generation}.
\newblock
\newblock
\showeprint[arxiv]{2309.06380}~[cs.LG]


\bibitem[Loshchilov and Hutter(2017)]%
        {loshchilov2017adamW}
\bibfield{author}{\bibinfo{person}{Ilya Loshchilov} {and} \bibinfo{person}{Frank Hutter}.} \bibinfo{year}{2017}\natexlab{}.
\newblock \showarticletitle{Decoupled weight decay regularization}.
\newblock \bibinfo{journal}{\emph{arXiv preprint arXiv:1711.05101}} (\bibinfo{year}{2017}).
\newblock


\bibitem[Matusik et~al\mbox{.}(2005)]%
        {simplicialComplex}
\bibfield{author}{\bibinfo{person}{Wojciech Matusik}, \bibinfo{person}{Matthias Zwicker}, {and} \bibinfo{person}{Fr{\'e}do Durand}.} \bibinfo{year}{2005}\natexlab{}.
\newblock \showarticletitle{Texture design using a simplicial complex of morphable textures}.
\newblock \bibinfo{journal}{\emph{ACM Transactions on Graphics (TOG)}} \bibinfo{volume}{24}, \bibinfo{number}{3} (\bibinfo{year}{2005}), \bibinfo{pages}{787--794}.
\newblock


\bibitem[Nichol and Dhariwal(2021)]%
        {nichol2021improvedDiffusion}
\bibfield{author}{\bibinfo{person}{Alexander~Quinn Nichol} {and} \bibinfo{person}{Prafulla Dhariwal}.} \bibinfo{year}{2021}\natexlab{}.
\newblock \showarticletitle{Improved denoising diffusion probabilistic models}. In \bibinfo{booktitle}{\emph{International Conference on Machine Learning}}. PMLR, \bibinfo{pages}{8162--8171}.
\newblock


\bibitem[Park et~al\mbox{.}(2019)]%
        {park2019SPADE}
\bibfield{author}{\bibinfo{person}{Taesung Park}, \bibinfo{person}{Ming-Yu Liu}, \bibinfo{person}{Ting-Chun Wang}, {and} \bibinfo{person}{Jun-Yan Zhu}.} \bibinfo{year}{2019}\natexlab{}.
\newblock \showarticletitle{Semantic Image Synthesis with Spatially-Adaptive Normalization}. In \bibinfo{booktitle}{\emph{Proceedings of the IEEE Conference on Computer Vision and Pattern Recognition}}.
\newblock


\bibitem[Perlin(1985)]%
        {perlin85}
\bibfield{author}{\bibinfo{person}{Ken Perlin}.} \bibinfo{year}{1985}\natexlab{}.
\newblock \showarticletitle{An Image Synthesizer}. In \bibinfo{booktitle}{\emph{Proceedings of the 12th Annual Conference on Computer Graphics and Interactive Techniques}} \emph{(\bibinfo{series}{SIGGRAPH '85})}. \bibinfo{publisher}{Association for Computing Machinery}, \bibinfo{address}{New York, NY, USA}, \bibinfo{pages}{287–296}.
\newblock
\showISBNx{0897911660}
\urldef\tempurl%
\url{https://doi.org/10.1145/325334.325247}
\showDOI{\tempurl}


\bibitem[Perlin(2002)]%
        {perlin2002improving}
\bibfield{author}{\bibinfo{person}{Ken Perlin}.} \bibinfo{year}{2002}\natexlab{}.
\newblock \showarticletitle{Improving noise}. In \bibinfo{booktitle}{\emph{Proceedings of the 29th annual conference on Computer graphics and interactive techniques}}. \bibinfo{pages}{681--682}.
\newblock


\bibitem[Raad et~al\mbox{.}(2018)]%
        {raad2018textureSurvey}
\bibfield{author}{\bibinfo{person}{Lara Raad}, \bibinfo{person}{Axel Davy}, \bibinfo{person}{Agn{\`e}s Desolneux}, {and} \bibinfo{person}{Jean-Michel Morel}.} \bibinfo{year}{2018}\natexlab{}.
\newblock \showarticletitle{A survey of exemplar-based texture synthesis}.
\newblock \bibinfo{journal}{\emph{Annals of Mathematical Sciences and Applications}} \bibinfo{volume}{3}, \bibinfo{number}{1} (\bibinfo{year}{2018}), \bibinfo{pages}{89--148}.
\newblock


\bibitem[Ren et~al\mbox{.}(2021)]%
        {ren2021csg}
\bibfield{author}{\bibinfo{person}{Daxuan Ren}, \bibinfo{person}{Jianmin Zheng}, \bibinfo{person}{Jianfei Cai}, \bibinfo{person}{Jiatong Li}, \bibinfo{person}{Haiyong Jiang}, \bibinfo{person}{Zhongang Cai}, \bibinfo{person}{Junzhe Zhang}, \bibinfo{person}{Liang Pan}, \bibinfo{person}{Mingyuan Zhang}, \bibinfo{person}{Haiyu Zhao}, {et~al\mbox{.}}} \bibinfo{year}{2021}\natexlab{}.
\newblock \showarticletitle{CSG-Stump: A Learning Friendly CSG-Like Representation for Interpretable Shape Parsing}. In \bibinfo{booktitle}{\emph{Proceedings of the IEEE/CVF International Conference on Computer Vision}}. \bibinfo{pages}{12478--12487}.
\newblock


\bibitem[Rodriguez-Pardo and Garces(2023)]%
        {SeamlessGAN}
\bibfield{author}{\bibinfo{person}{Carlos Rodriguez-Pardo} {and} \bibinfo{person}{Elena Garces}.} \bibinfo{year}{2023}\natexlab{}.
\newblock \showarticletitle{SeamlessGAN: Self-Supervised Synthesis of Tileable Texture Maps}.
\newblock \bibinfo{journal}{\emph{IEEE Transactions on Visualization and Computer Graphics}} \bibinfo{volume}{29}, \bibinfo{number}{6} (\bibinfo{date}{jun} \bibinfo{year}{2023}), \bibinfo{pages}{2914–2925}.
\newblock
\showISSN{1077-2626}
\urldef\tempurl%
\url{https://doi.org/10.1109/TVCG.2022.3143615}
\showDOI{\tempurl}


\bibitem[Schonfeld et~al\mbox{.}(2020)]%
        {schonfeldCutMixApplication}
\bibfield{author}{\bibinfo{person}{Edgar Schonfeld}, \bibinfo{person}{Bernt Schiele}, {and} \bibinfo{person}{Anna Khoreva}.} \bibinfo{year}{2020}\natexlab{}.
\newblock \showarticletitle{A u-net based discriminator for generative adversarial networks}. In \bibinfo{booktitle}{\emph{Proceedings of the IEEE/CVF conference on computer vision and pattern recognition}}. \bibinfo{pages}{8207--8216}.
\newblock


\bibitem[Sendik and Cohen-Or(2017)]%
        {DeepCorrelations}
\bibfield{author}{\bibinfo{person}{Omry Sendik} {and} \bibinfo{person}{Daniel Cohen-Or}.} \bibinfo{year}{2017}\natexlab{}.
\newblock \showarticletitle{Deep Correlations for Texture Synthesis}.
\newblock \bibinfo{journal}{\emph{ACM Trans. Graph.}} \bibinfo{volume}{36}, \bibinfo{number}{5}, Article \bibinfo{articleno}{161} (\bibinfo{date}{jul} \bibinfo{year}{2017}), \bibinfo{numpages}{15}~pages.
\newblock
\showISSN{0730-0301}
\urldef\tempurl%
\url{https://doi.org/10.1145/3015461}
\showDOI{\tempurl}


\bibitem[Shen et~al\mbox{.}(2021)]%
        {linearAttention}
\bibfield{author}{\bibinfo{person}{Zhuoran Shen}, \bibinfo{person}{Mingyuan Zhang}, \bibinfo{person}{Haiyu Zhao}, \bibinfo{person}{Shuai Yi}, {and} \bibinfo{person}{Hongsheng Li}.} \bibinfo{year}{2021}\natexlab{}.
\newblock \showarticletitle{Efficient attention: Attention with linear complexities}. In \bibinfo{booktitle}{\emph{Proceedings of the IEEE/CVF winter conference on applications of computer vision}}. \bibinfo{pages}{3531--3539}.
\newblock


\bibitem[Shi et~al\mbox{.}(2020)]%
        {Shi2020MATch}
\bibfield{author}{\bibinfo{person}{Liang Shi}, \bibinfo{person}{Beichen Li}, \bibinfo{person}{Milo{\v s} Ha{\v s}an}, \bibinfo{person}{Kalyan Sunkavalli}, \bibinfo{person}{Tamy Boubekeur}, \bibinfo{person}{Radomir Mech}, {and} \bibinfo{person}{Wojciech Matusik}.} \bibinfo{year}{2020}\natexlab{}.
\newblock \showarticletitle{MATch: Differentiable Material Graphs for Procedural Material Capture}.
\newblock \bibinfo{journal}{\emph{ACM Trans. Graph.}} \bibinfo{volume}{39}, \bibinfo{number}{6} (\bibinfo{date}{Dec.} \bibinfo{year}{2020}), \bibinfo{pages}{1--15}.
\newblock


\bibitem[Shoemake(1985)]%
        {SLERP}
\bibfield{author}{\bibinfo{person}{Ken Shoemake}.} \bibinfo{year}{1985}\natexlab{}.
\newblock \showarticletitle{Animating rotation with quaternion curves}. In \bibinfo{booktitle}{\emph{Proceedings of the 12th annual conference on Computer graphics and interactive techniques}}. \bibinfo{pages}{245--254}.
\newblock


\bibitem[Song et~al\mbox{.}(2020)]%
        {song2020ddim}
\bibfield{author}{\bibinfo{person}{Jiaming Song}, \bibinfo{person}{Chenlin Meng}, {and} \bibinfo{person}{Stefano Ermon}.} \bibinfo{year}{2020}\natexlab{}.
\newblock \showarticletitle{Denoising Diffusion Implicit Models}.
\newblock \bibinfo{journal}{\emph{arXiv:2010.02502}} (\bibinfo{date}{October} \bibinfo{year}{2020}).
\newblock
\urldef\tempurl%
\url{https://arxiv.org/abs/2010.02502}
\showURL{%
\tempurl}


\bibitem[Song(2021)]%
        {yangSongDiffusionBlog}
\bibfield{author}{\bibinfo{person}{Yang Song}.} \bibinfo{year}{2021}\natexlab{}.
\newblock \bibinfo{title}{Generative Modeling by Estimating Gradients of the Data Distribution}.
\newblock \bibinfo{howpublished}{\url{https://yang-song.net/blog/2021/score/}}.
\newblock


\bibitem[Tricard et~al\mbox{.}(2019)]%
        {PhasorNoise}
\bibfield{author}{\bibinfo{person}{Thibault Tricard}, \bibinfo{person}{Semyon Efremov}, \bibinfo{person}{C{\'e}dric Zanni}, \bibinfo{person}{Fabrice Neyret}, \bibinfo{person}{Jon{\`a}s Mart{\'\i}nez}, {and} \bibinfo{person}{Sylvain Lefebvre}.} \bibinfo{year}{2019}\natexlab{}.
\newblock \showarticletitle{Procedural phasor noise}.
\newblock \bibinfo{journal}{\emph{ACM Transactions on Graphics (TOG)}} \bibinfo{volume}{38}, \bibinfo{number}{4} (\bibinfo{year}{2019}), \bibinfo{pages}{1--13}.
\newblock


\bibitem[Ulyanov et~al\mbox{.}(2016)]%
        {TextureNetworks}
\bibfield{author}{\bibinfo{person}{Dmitry Ulyanov}, \bibinfo{person}{Vadim Lebedev}, \bibinfo{person}{Vedaldi Andrea}, {and} \bibinfo{person}{Victor Lempitsky}.} \bibinfo{year}{2016}\natexlab{}.
\newblock \showarticletitle{Texture Networks: Feed-forward Synthesis of Textures and Stylized Images}. In \bibinfo{booktitle}{\emph{Proceedings of The 33rd International Conference on Machine Learning}} \emph{(\bibinfo{series}{Proceedings of Machine Learning Research}, Vol.~\bibinfo{volume}{48})}, \bibfield{editor}{\bibinfo{person}{Maria~Florina Balcan} {and} \bibinfo{person}{Kilian~Q. Weinberger}} (Eds.). \bibinfo{publisher}{PMLR}, \bibinfo{address}{New York, New York, USA}, \bibinfo{pages}{1349--1357}.
\newblock


\bibitem[Um et~al\mbox{.}(2024)]%
        {diffusionLowDensity}
\bibfield{author}{\bibinfo{person}{Soobin Um}, \bibinfo{person}{Suhyeon Lee}, {and} \bibinfo{person}{Jong~Chul Ye}.} \bibinfo{year}{2024}\natexlab{}.
\newblock \showarticletitle{Don't Play Favorites: Minority Guidance for Diffusion Models}. In \bibinfo{booktitle}{\emph{The Twelfth International Conference on Learning Representations}}.
\newblock
\urldef\tempurl%
\url{https://openreview.net/forum?id=3NmO9lY4Jn}
\showURL{%
\tempurl}


\bibitem[Van Der~Linden et~al\mbox{.}(2013)]%
        {van2013proceduralDungeons}
\bibfield{author}{\bibinfo{person}{Roland Van Der~Linden}, \bibinfo{person}{Ricardo Lopes}, {and} \bibinfo{person}{Rafael Bidarra}.} \bibinfo{year}{2013}\natexlab{}.
\newblock \showarticletitle{Procedural generation of dungeons}.
\newblock \bibinfo{journal}{\emph{IEEE Transactions on Computational Intelligence and AI in Games}} \bibinfo{volume}{6}, \bibinfo{number}{1} (\bibinfo{year}{2013}), \bibinfo{pages}{78--89}.
\newblock


\bibitem[Worley(1996)]%
        {worleyNoise}
\bibfield{author}{\bibinfo{person}{Steven Worley}.} \bibinfo{year}{1996}\natexlab{}.
\newblock \showarticletitle{A cellular texture basis function}. In \bibinfo{booktitle}{\emph{Proceedings of the 23rd Annual Conference on Computer Graphics and Interactive Techniques}} \emph{(\bibinfo{series}{SIGGRAPH '96})}. \bibinfo{publisher}{Association for Computing Machinery}, \bibinfo{address}{New York, NY, USA}, \bibinfo{pages}{291–294}.
\newblock
\showISBNx{0897917464}
\urldef\tempurl%
\url{https://doi.org/10.1145/237170.237267}
\showDOI{\tempurl}


\bibitem[Wu and He(2018)]%
        {group_norm}
\bibfield{author}{\bibinfo{person}{Yuxin Wu} {and} \bibinfo{person}{Kaiming He}.} \bibinfo{year}{2018}\natexlab{}.
\newblock \showarticletitle{Group Normalization}. In \bibinfo{booktitle}{\emph{Computer Vision - {ECCV} 2018 - 15th European Conference, Munich, Germany, September 8-14, 2018, Proceedings, Part {XIII}}} \emph{(\bibinfo{series}{Lecture Notes in Computer Science}, Vol.~\bibinfo{volume}{11217})}, \bibfield{editor}{\bibinfo{person}{Vittorio Ferrari}, \bibinfo{person}{Martial Hebert}, \bibinfo{person}{Cristian Sminchisescu}, {and} \bibinfo{person}{Yair Weiss}} (Eds.). \bibinfo{publisher}{Springer}, \bibinfo{pages}{3--19}.
\newblock
\urldef\tempurl%
\url{https://doi.org/10.1007/978-3-030-01261-8\_1}
\showDOI{\tempurl}


\bibitem[Xiao et~al\mbox{.}(2022)]%
        {xiao2022tackling}
\bibfield{author}{\bibinfo{person}{Zhisheng Xiao}, \bibinfo{person}{Karsten Kreis}, {and} \bibinfo{person}{Arash Vahdat}.} \bibinfo{year}{2022}\natexlab{}.
\newblock \showarticletitle{Tackling the Generative Learning Trilemma with Denoising Diffusion GANs}. In \bibinfo{booktitle}{\emph{ICLR}}.
\newblock


\bibitem[Yun et~al\mbox{.}(2019)]%
        {yun2019cutmix}
\bibfield{author}{\bibinfo{person}{Sangdoo Yun}, \bibinfo{person}{Dongyoon Han}, \bibinfo{person}{Seong~Joon Oh}, \bibinfo{person}{Sanghyuk Chun}, \bibinfo{person}{Junsuk Choe}, {and} \bibinfo{person}{Youngjoon Yoo}.} \bibinfo{year}{2019}\natexlab{}.
\newblock \showarticletitle{CutMix: Regularization Strategy to Train Strong Classifiers with Localizable Features}. In \bibinfo{booktitle}{\emph{International Conference on Computer Vision (ICCV)}}.
\newblock


\bibitem[Zhang et~al\mbox{.}(2018)]%
        {LPIPS}
\bibfield{author}{\bibinfo{person}{Richard Zhang}, \bibinfo{person}{Phillip Isola}, \bibinfo{person}{Alexei~A Efros}, \bibinfo{person}{Eli Shechtman}, {and} \bibinfo{person}{Oliver Wang}.} \bibinfo{year}{2018}\natexlab{}.
\newblock \showarticletitle{The Unreasonable Effectiveness of Deep Features as a Perceptual Metric}. In \bibinfo{booktitle}{\emph{CVPR}}.
\newblock


\bibitem[Zhou et~al\mbox{.}(2023b)]%
        {zhou2023photomat}
\bibfield{author}{\bibinfo{person}{Xilong Zhou}, \bibinfo{person}{Milos Hasan}, \bibinfo{person}{Valentin Deschaintre}, \bibinfo{person}{Paul Guerrero}, \bibinfo{person}{Yannick Hold-Geoffroy}, \bibinfo{person}{Kalyan Sunkavalli}, {and} \bibinfo{person}{Nima~Khademi Kalantari}.} \bibinfo{year}{2023}\natexlab{b}.
\newblock \showarticletitle{PhotoMat: A Material Generator Learned from Single Flash Photos}. In \bibinfo{booktitle}{\emph{ACM SIGGRAPH 2023 Conference Proceedings}}. \bibinfo{pages}{1--11}.
\newblock


\bibitem[Zhou et~al\mbox{.}(2023a)]%
        {TexSynGuidedCorrespondence}
\bibfield{author}{\bibinfo{person}{Yang Zhou}, \bibinfo{person}{Kaijian Chen}, \bibinfo{person}{Rongjun Xiao}, {and} \bibinfo{person}{Hui Huang}.} \bibinfo{year}{2023}\natexlab{a}.
\newblock \showarticletitle{Neural Texture Synthesis with Guided Correspondence}. In \bibinfo{booktitle}{\emph{Conference on Computer Vision and Pattern Recognition (CVPR)}}. \bibinfo{pages}{18095--18104}.
\newblock


\bibitem[Zhou et~al\mbox{.}(2018)]%
        {TexSynAdversarialExpansion}
\bibfield{author}{\bibinfo{person}{Yang Zhou}, \bibinfo{person}{Zhen Zhu}, \bibinfo{person}{Xiang Bai}, \bibinfo{person}{Dani Lischinski}, \bibinfo{person}{Daniel Cohen-Or}, {and} \bibinfo{person}{Hui Huang}.} \bibinfo{year}{2018}\natexlab{}.
\newblock \showarticletitle{Non-stationary Texture Synthesis by Adversarial Expansion}.
\newblock \bibinfo{journal}{\emph{ACM Transactions on Graphics (Proc. SIGGRAPH)}} \bibinfo{volume}{37}, \bibinfo{number}{4} (\bibinfo{year}{2018}), \bibinfo{pages}{49:1--49:13}.
\newblock


\end{thebibliography}

\clearpage
\appendix
\definecolor{lightgray}{gray}{0.9}
\begin{table*}
    \centering
    \begin{tabular}{c|c|c}
        \textbf{Noise function} & \textbf{Sampled parameters} & \textbf{Range} \\
        \hline
        \rowcolor{lightgray}
        {cells 4} & {scale} & $i[5,50]$ \\
        
        \rowcolor{white}
        {cells 1} & {scale} & $i[10,50]$ \\

        \rowcolor{lightgray} {voronoi} & {scale} & $f[5.0,15.0]$ \\
        \rowcolor{lightgray} & {distortion intensity} & $f[0.0,1.0]$ \\
        \rowcolor{lightgray} & {distortion scale multiplier} & $f[1.0,2.0]$ \\

        \rowcolor{white}
        {microscope view} & {scale} & $i[25,64]$ \\
        & {warp intensity} & $f[0.0,1.0]$ \\
        
        \rowcolor{lightgray} {bnw spots 1} & {scale} & $i[1,3]$ \\
        
        {liquid} & {scale} & $i[10,45]$ \\
        & {warp intensity} & $f[0.1,0.8]$ \\
        
        \rowcolor{lightgray} {grunge galvanic small} & {crispness} & $f[0.0,0.75]$ \\
        \rowcolor{lightgray} & {dirt} & $f[0.0,1.0]$ \\
        \rowcolor{lightgray} & {micro distortion} & $f[0.0,0.6]$ \\
        
        {grunge leaky paint} & {leak intensity} & $f[0.0,1.0]$ \\
        & {leak scale} & $i[1,8]$ \\
        & {leak crispness} & $f[0.0,0.8]$ \\
                
    \end{tabular}
    \begin{tabular}{c|c|c}
        \textbf{Noise function} & \textbf{Sampled parameters} & \textbf{Range} \\
        \hline        
        \rowcolor{lightgray} {grunge rust fine} & {base grunge contrast} & $f[-0.3,0.3]$ \\
        \rowcolor{lightgray} & {base warp intensity} & $f[0.0,256.0]$ \\
        
        {grunge damas} & {distortion} & $f[0.0,1.0]$ \\
        & {divisions} & $i[4,16]$ \\
        & {waves} & $i[1,3]$ \\
        & {details} & $f[0.0,0.75]$ \\
        & {rotation random} & $f[0.05,1.0]$ \\
        
        \rowcolor{lightgray} {grunge map 002} & N/A & N/A \\
        
        {grunge map 005} & N/A & N/A \\
        
        \rowcolor{lightgray} {messy fibers 3} & {scale} & $i[1,3]$ \\
        
        {perlin} & {scale} & $i[10,50]$ \\
        
        \rowcolor{lightgray} {gaussian} & {scale} & $i[10,50]$ \\
        
        {clouds 1} & {scale} & $i[1,5]$ \\
        \rowcolor{lightgray} {clouds 2} & {scale} & $i[1,5]$ \\
        {clouds 3} & {scale} & $i[1,3]$ \\

       \rowcolor{lightgray} & & \\
        
    \end{tabular}
    \caption{We enumerate the noise functions that are sampled from Adobe Substance 3D Designer, along with their parameters and accompanying ranges -- integer and real ranges are denoted by $i$ and $f$ respectively.}
    \label{tab:noiseFunctions}
\end{table*}

\section{Experimental and Implementation Details}

\begin{table}
    \centering
    \begin{tabular}{c|c|c|c|c|c}
     & \multicolumn{2}{c|}{\textbf{FID}$\downarrow$} & & \multicolumn{2}{c}{\textbf{FID}$\downarrow$} \\
    \hline
    \textbf{Noise} & \textbf{PSGAN} & \textbf{Ours} & \textbf{Noise} & \textbf{PSGAN} & \textbf{Ours} \\
    \hline
    \rowcolor{lightgray}
    cells 4 & $218.8$ & $33.6$ & rust fine & $88.5$ & $12.3$ \\
    \rowcolor{white}
    cells 1 & $171.3$ & $2.4$ & damas & $56.1$ & $71.0$ \\
    \rowcolor{lightgray}
    voronoi & $149.3$ & $12.5$ & map 002 & $37.3$ & $13.7$ \\
    \rowcolor{white}
    microscope & $133.1$ & $29.7$ & map 005 & $155.4$ & $34.5$ \\
    \rowcolor{lightgray}
    bnw spots 1 & $22.4$ & $4.4$ & fibers & $86.4$ & $34.6$ \\
    \rowcolor{white}
    liquid & $163.7$ & $38.0$ & perlin & $47.8$ & $4.8$ \\
    \rowcolor{lightgray}
    galvanic small & $79.2$ & $24.9$ & gaussian & $45.2$ & $1.7$ \\
    \rowcolor{white}
    leaky paint & $155.1$ & $44.3$ & clouds 1 & $20.4$ & $1.5$ \\
    \rowcolor{lightgray}    
    clouds 3 & $38.4$ & $2.9$ & clouds 2 & $110.8$ & $9.0$ \\
    \hline
    \textbf{Mean} & $99.2$ & $20.9$ & \textbf{Median} & $87.5$ & $13.1$ \\
    \hline
    \end{tabular}
    \caption{We compare our FID scores for each noise type alongside PSGAN. The mean and median of all values are shown as well.}
    \label{tab:fidMain}
\end{table}

\subsection{Noise dataset details}
We include a table enumerating all sampled noise types, the parameters that we sample, as well as the parameter ranges (see Table \ref{tab:noiseFunctions}). In general, we choose to include any parameters that lead to noticeable changes in the resulting noise, with exceptions to parameters that simply act as color correction (i.e. \texttt{grunge map 005}'s \emph{contrast} parameter). We note that, despite their names, some of these noises correspond to those that belong to graphics literature; for instance, \texttt{cells 4}, \texttt{cells 1}, and \texttt{voronoi} are variants of Worley noise \cite{worleyNoise}. For more details about these noises, please refer to the Adobe Substance documentation~\shortcite{substanceDocumentation}. 

\begin{figure*}[p]
    \centering
    \includegraphics[width=\textwidth]{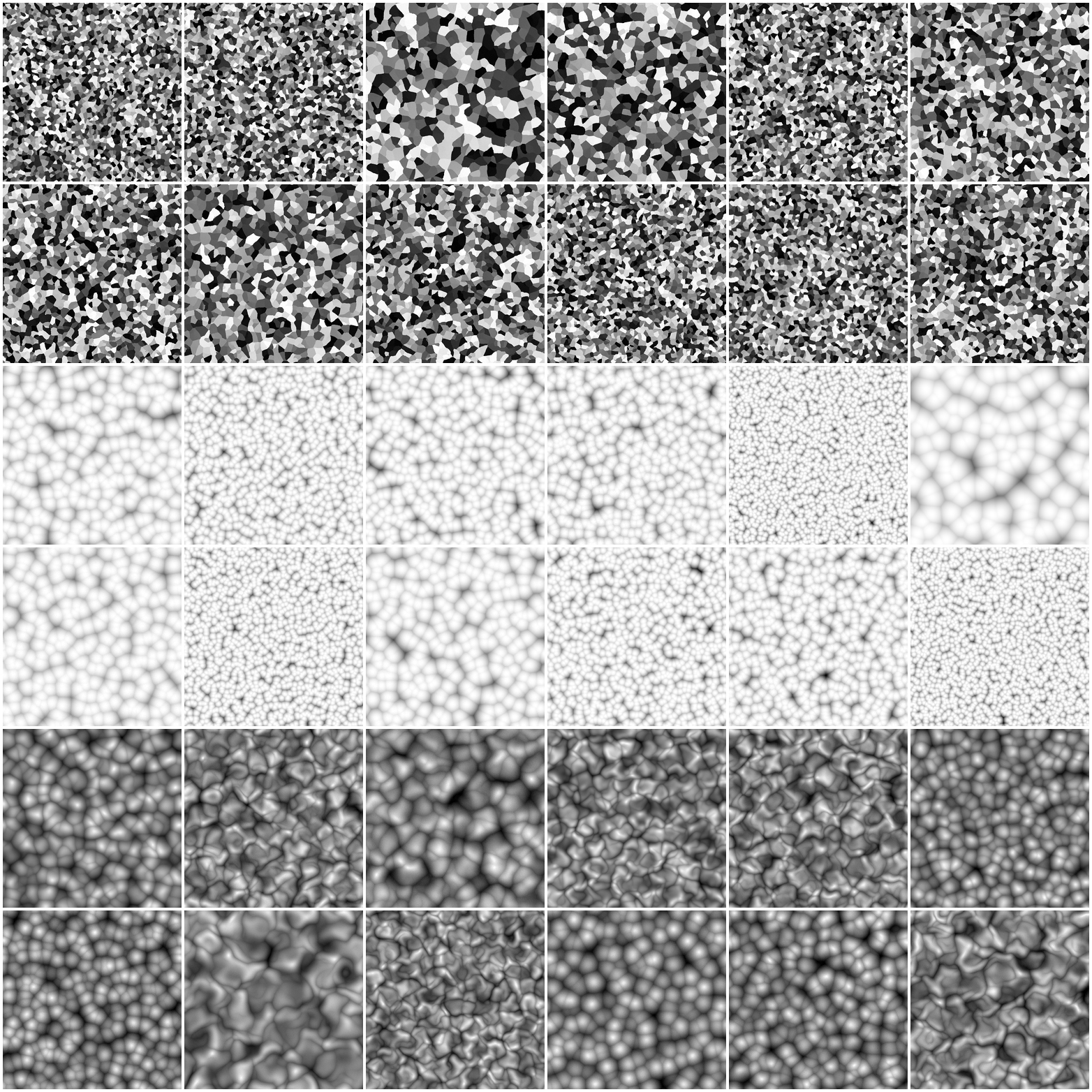}
    \caption{Random samples of our model's \texttt{cells 4}, \texttt{cells 1}, and \texttt{voronoi} noises at $256\times256$ resolution.}
    \label{fig:grid0}
\end{figure*}

\begin{figure*}[p]
    \centering
    \includegraphics[width=\textwidth]{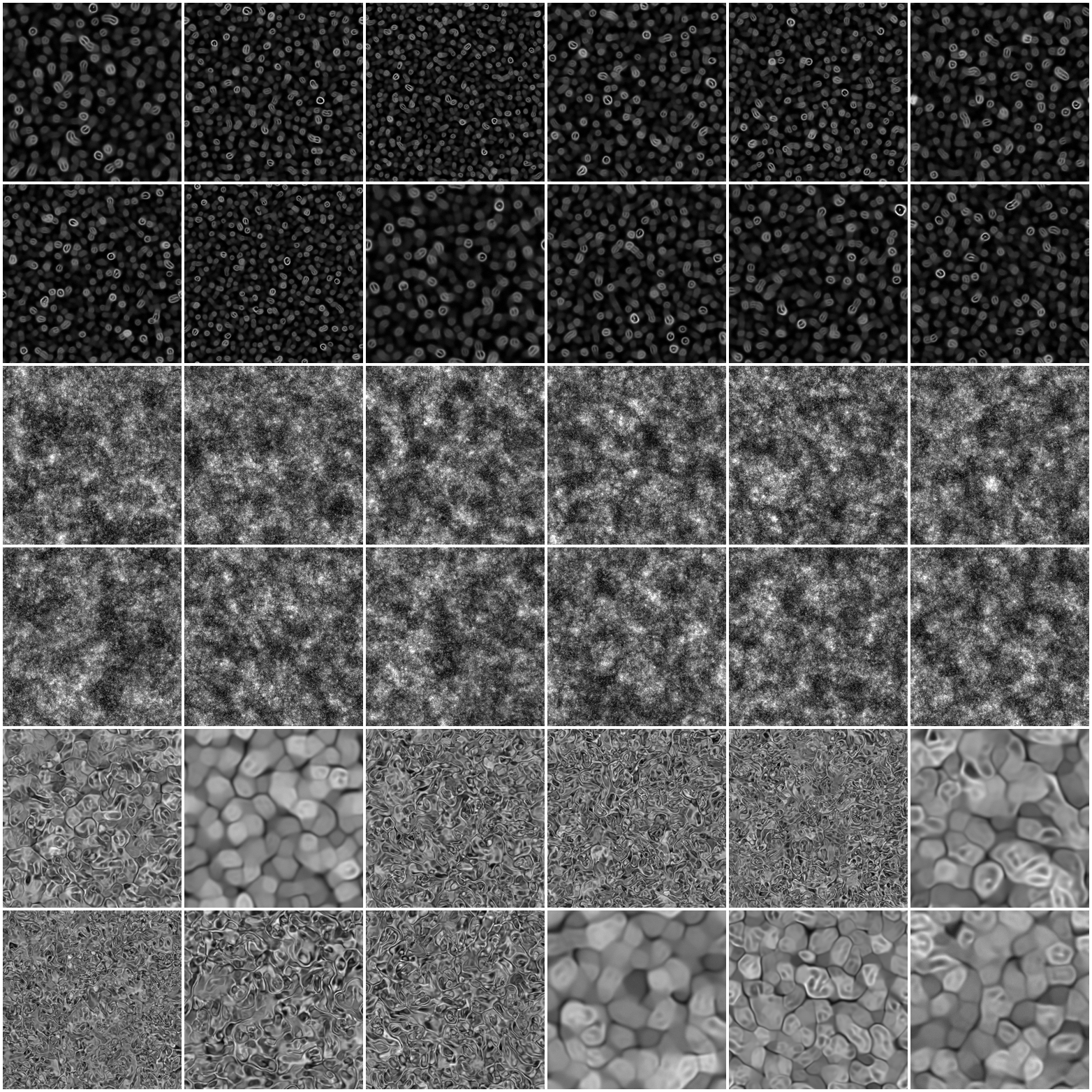}
    \caption{Random samples of our model's \texttt{microscope view}, \texttt{bnw spots1}, and \texttt{liquid} noises at $256\times256$ resolution.}
    \label{fig:grid1}
\end{figure*}

\begin{figure*}[p]
    \centering
    \includegraphics[width=\textwidth]{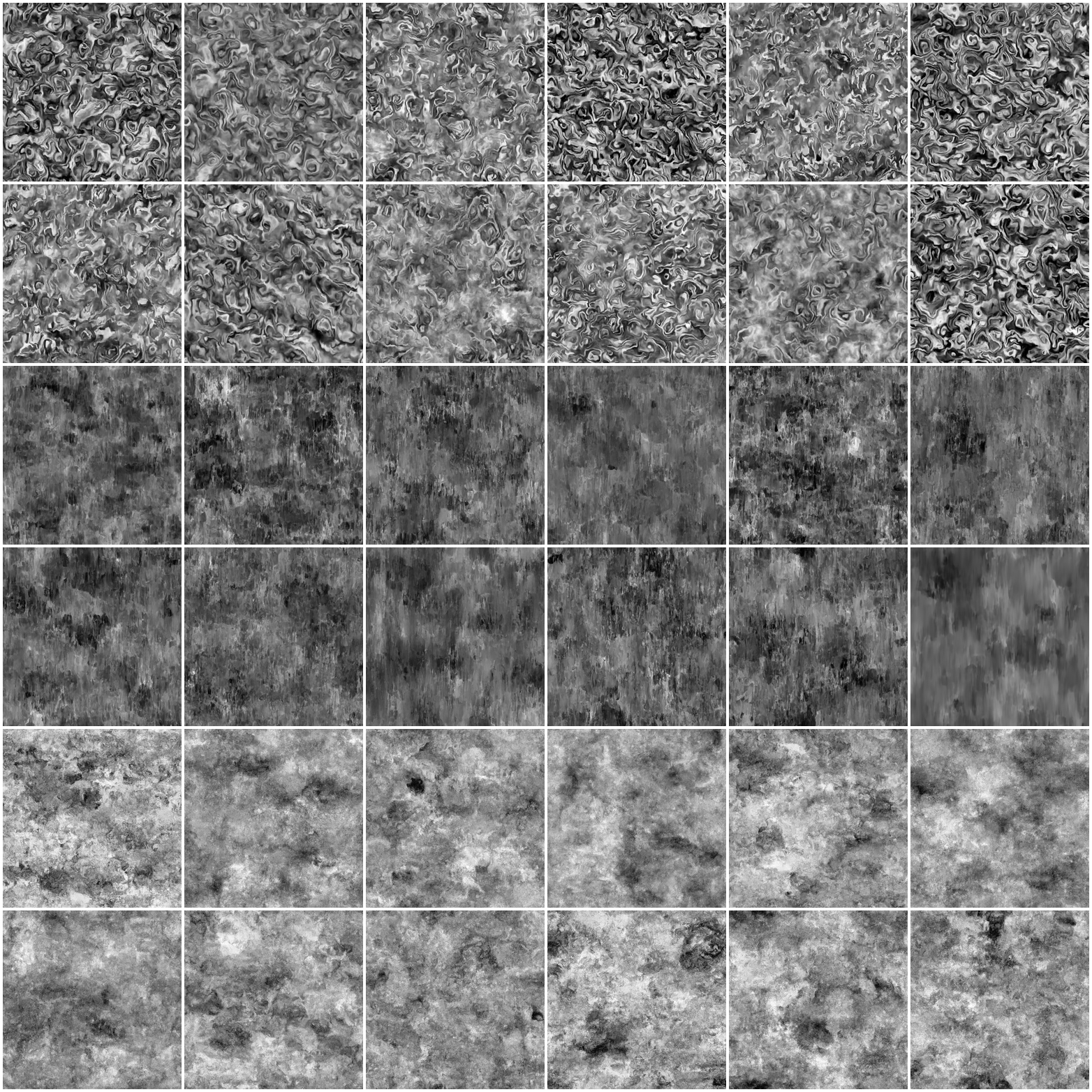}
    \caption{Random samples of our model's \texttt{grunge galvanic small}, \texttt{grunge leaky paint}, and \texttt{grunge rust fine} noises at $256\times256$ resolution.}
    \label{fig:grid2}
\end{figure*}

\begin{figure*}[p]
    \centering
    \includegraphics[width=\textwidth]{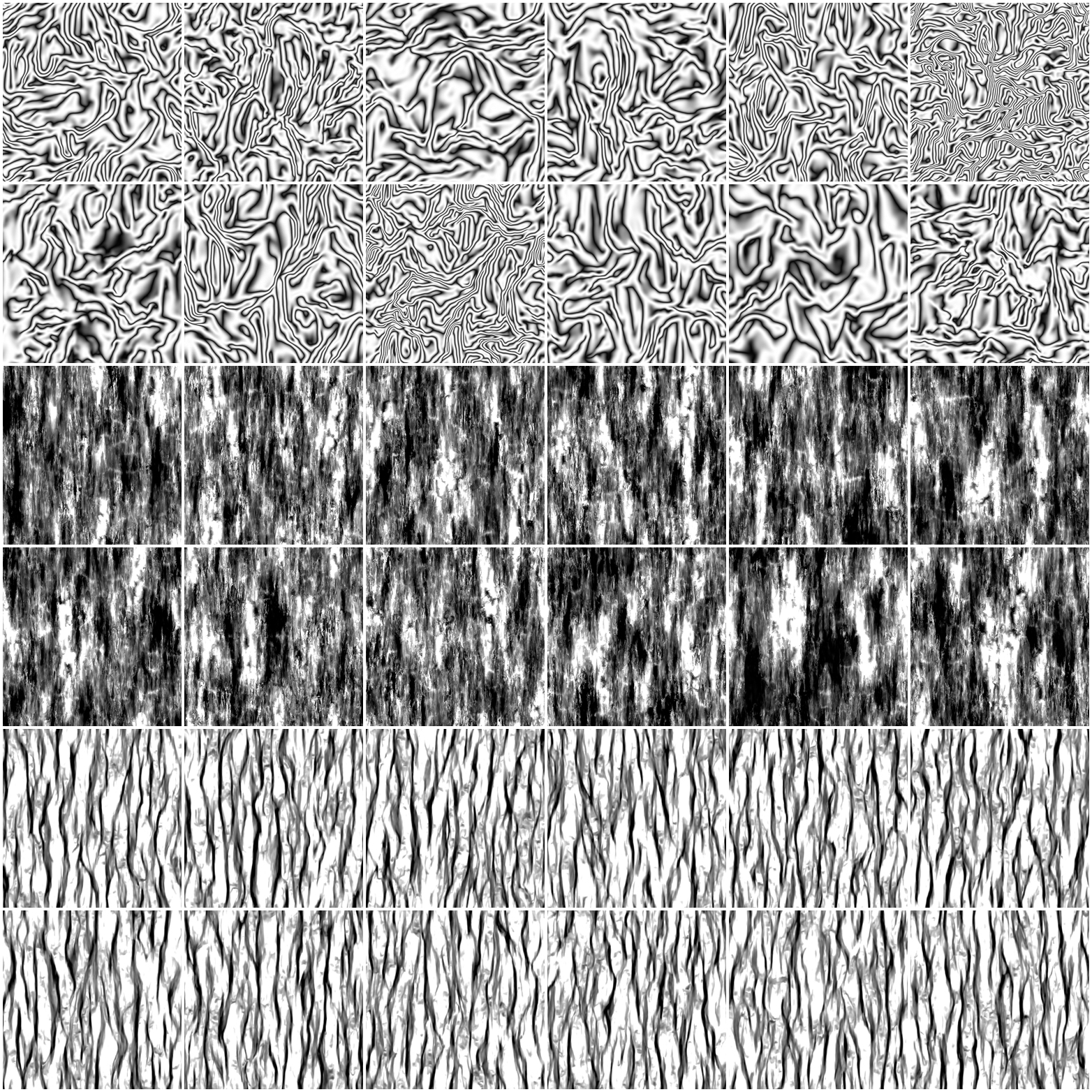}
    \caption{Random samples of our model's \texttt{grunge damas}, \texttt{grunge map 002}, and \texttt{grunge map 005} noises at $256\times256$ resolution.}
    \label{fig:grid3}
\end{figure*}

\begin{figure*}[p]
    \centering
    \includegraphics[width=\textwidth]{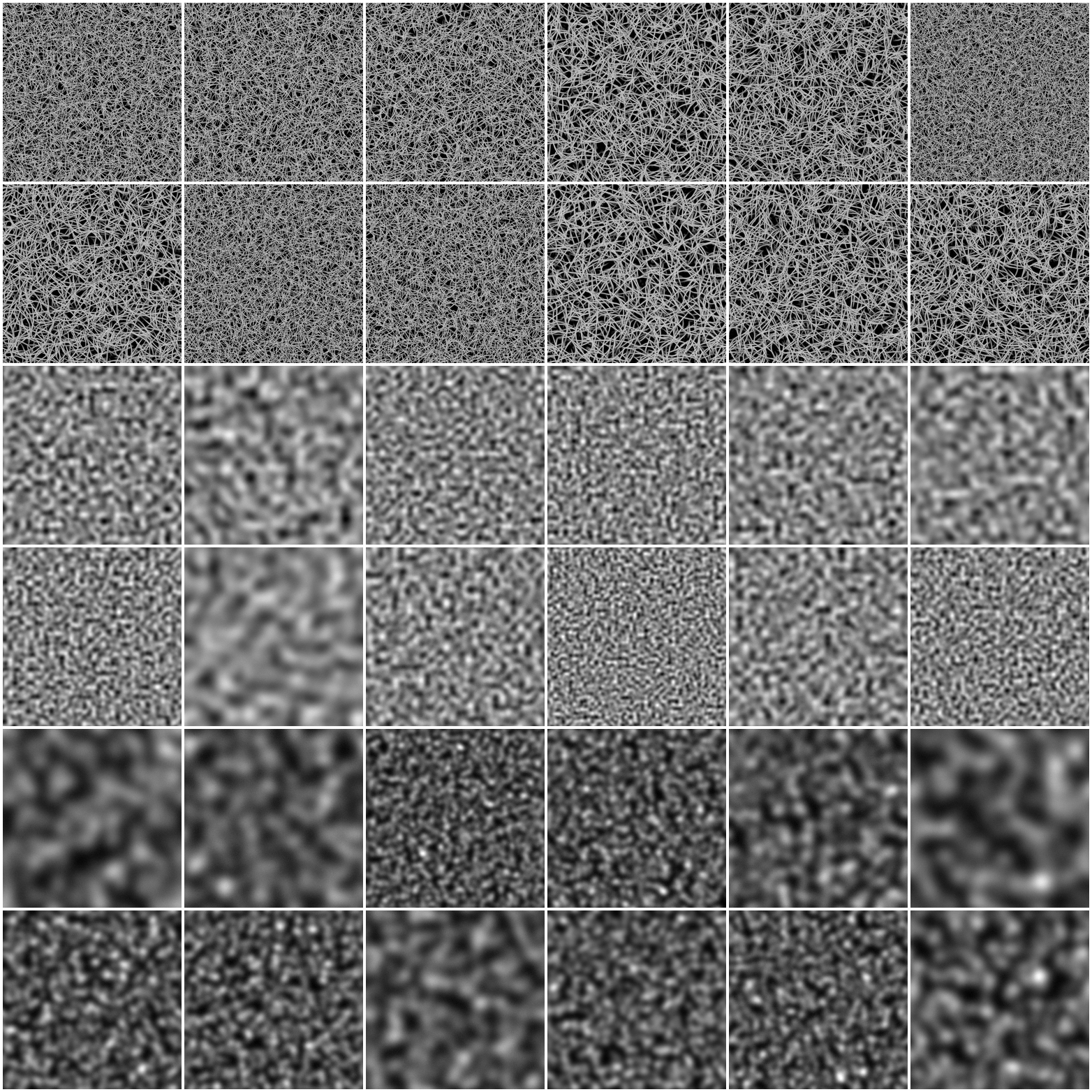}
    \caption{Random samples of our model's \texttt{messy fibers 3}, \texttt{perlin}, and \texttt{gaussian} noises at $256\times256$ resolution.}
    \label{fig:grid4}
\end{figure*}

The conditioning vector $\*f_p$ contains an entry for each unique noise parameter -- we treat identical parameter names as separate, with the exception of \texttt{scale}, which is treated as a single entry in the vector. Parameters are independently normalized to the range $[0,1]$ before being passed to our conditioning MLP. Below we show many samples of our model's outputs with all listed noise parameters being sampled randomly (see \cref{fig:grid0,fig:grid1,fig:grid2,fig:grid3,fig:grid4}).

\begin{figure}
    \centering
    \includegraphics[width=\linewidth]{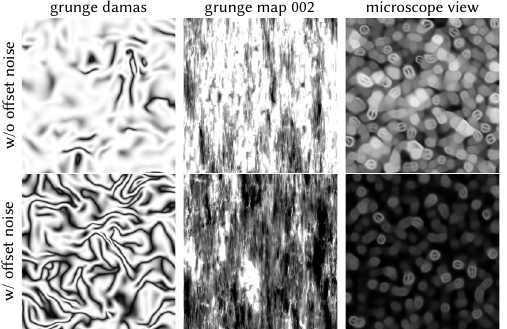}
    \caption{Ablation of offset noise. Without offset noise, the network occasionally fails to synthesize noise images with extreme intensity distributions (i.e. intensely dark and bright images). We show representative examples above.}
    \label{fig:ablationOffsetNoise}
\end{figure}

\subsection{PSGAN Baseline}
We slightly modify the PSGAN~\shortcite{PeriodicSpatialGAN} architecture for the results shown in Figure \ref{fig:oursPSGANImageMeld}. Since PSGAN is a purely unconditional generative model (i.e. it has no class or parameter conditioning beyond randomized latent code inputs), we endow the PSGAN model with SPADE conditioning blocks that are identical to the ones used throughout our method. We made the discriminator slightly larger to compensate for the added parameters in the generator. Finally, we use the WGAN loss \cite{WGAN}. The remaining parts of the network and training are largely kept as-is from the open implementation. In total the generator has 30 million parameters.

\subsection{Inverse Material Design Details}\label{supp:InverseDesign}
We expose three parameters to the MATch differentiable material graph optimizer; a soft-class vector $\*v_c$, the parameter vector $\*f_p$, and the diffusion Gaussian noise image $\*z$. The soft-class vector represents a list of $[0,1]$ values that are used to index into our class embeddings -- this amounts to taking a convex combination over the class embeddings. During optimization, we apply the SoftMax function to $\*v_c$ to ensure the values are valid. In order to encourage sparsity, we include an L1 regularization term on both $\*v_c$ and $\*f_p$, using a weight of $0.1$ for both L1 terms in the final optimization objective. Additionally, we include a scheduled temperature parameter $\tau$ into the SoftMax operator by performing element-wise division of $\*v_c$ by $\tau$, $\text{softmax}(v, t) = \frac{e^{v_i/t}}{\sum_{j=1}^{K} e^{v_j/t}}$.
The temperature is initialized to $0.25$ and is updated at every optimization step via the update rule $\tau' \leftarrow \tau \cdot 0.97$. Finally we clamp $\tau$ to be at minimum $0.01$. This scheduled temperature forces the optimization to ``hone in'' on a single class. We use a learning rate of $0.01$ for all graphs. After $\tau = 0.01$ we perform a warm-restart of the optimizer and add noise to the exposed parameters to avoid local minima.

\section{Training with additional noise types}\label{sec:supp_additionalNoise} 
\begin{figure}
    \centering
    \includegraphics{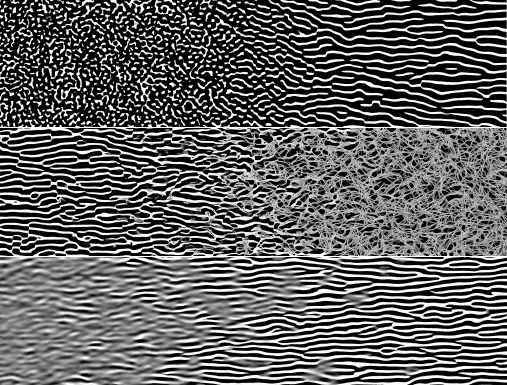}
    \caption{Examples of our model's Phasor noise. From top to bottom: isotropic Phasor to anisotropic Phasor (parameter interpolation), anisotropic Phasor to Messy Fibers, and anisotropic Gabor to anisotropic Phasor (class interpolation).}
    \label{fig:phasorNoise}
\end{figure}

\begin{figure}
    \centering
    \includegraphics{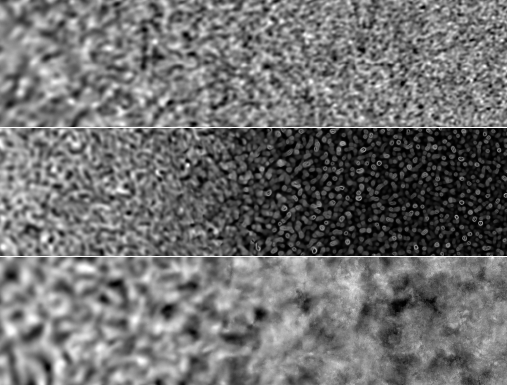}
    \caption{Examples of our model's Gabor noise. From top to bottom: Gabor to Gabor (parameter interpolation), Gabor to Microscope View, and Gabor to Grunge Rust Fine (class interpolation).}
    \label{fig:gaborNoise}
\end{figure}

We train our model on two additional noise types that are commonly used in graphics literature: Phasor noise~\shortcite{PhasorNoise} and Gabor noise~\shortcite{lagae2009gabor}. We utilize the same data sampling method as mentioned in Section \ref{sec:implementation}. The parameters that we sampled for both noise types, as well as their ranges, are enumerated in Table \ref{tab:additionalNoise}. We use the released implementations for both noise types to accrue our training data. Examples of spatially-varying images using parameter interpolation and class interpolation for both Phasor and Gabor noise are shown in \cref{fig:phasorNoise,fig:gaborNoise}.

We note that our model exhibits an FID of 93.6 and 129.3 on Phasor and Gabor noise respectively, which is notably higher than the FID scores for other noise types. Upon inspection of our model's outputs, we see that some parameter configurations are not well captured by our training. For example, Gabor noise exhibits a significantly different visual appearance when its principal frequency and kernel width are very small; however, since this appearance is only captured by a narrow subset of the parameter space, the dataset thus contains much fewer of such samples. One drawback of diffusion models is that they model low-density regions of the data distribution less accurately (compared to higher-density regions) ~\cite{yangSongDiffusionBlog,diffusionLowDensity}, and hence our performance is worse in such situations. More careful data sampling methods may be needed when the desired noise distribution exhibits low-density modes; e.g. by sampling some regions of the parameter space more frequently.

\begin{table}
\begin{tabular}{c|c|c}
    \textbf{Noise function} & \textbf{Sampled parameters} & \textbf{Range} \\
    \hline
    {phasor noise} & {principal frequency, $F$} & $f[8.0,32.0]$ \\
    & {num cells} & $i[1,12]$ \\
    & {phasor density} & $f[0.3,0.5]$ \\
    & {factor angle spread, $\theta$} & $f[0.0,1.0]$ \\

    \rowcolor{lightgray} {gabor noise} & {principal frequency, $F$} & $f[0.02,0.08]$ \\
    \rowcolor{lightgray} & {kernel width, $\alpha$} & $f[0.01,0.35]$ \\
    \rowcolor{lightgray} & {kernel orientation, $\omega$} & $f[0.0,2\pi]$ \\
\end{tabular}
\caption{Parameters and sampling ranges for additional noise types.}
\label{tab:additionalNoise}
\end{table}

\section{Model architectures}\label{sec:supp_Architectures}
The U-Net model detailed in Section \ref{sec:implementation} has just ${\sim}5.1$ million parameters and is constructed using two downsampling blocks, a bottleneck block, and two upsampling blocks. Each block contains two ResNet sub-blocks. An additional ResNet sub-block is placed at the end of the network. We use channel dimensions of 32 and 64 for the outer blocks, and 128 for the bottleneck. The ResNet sub-blocks are globally conditioned on the diffusion time $t$, and spatially conditioned (via SPADE) by 128-dim noise embeddings. We will refer to this model as \texttt{Model-XS} (extra small) below.

We train two additional model architectures to evaluate the potential performance of larger and more complex models. \texttt{Model-S} is identical to \texttt{Model-XS}, with the exception of added \emph{linear attention} layers inside of each block \cite{linearAttention}, and the addition of an extra set of downsampling/upsampling blocks. Similarly, \texttt{Model-M} is identical to \texttt{Model-S}, with the exception of larger channel dimensions. In Table \ref{tab:modelArchs} we summarize these architectures as well as their FID scores and inference performance. The models were trained for the same number of optimization steps following the details in Section \ref{sec:implementation}.

For our inverse material design application, we found that using \texttt{Model-XS} was most suitable due to the added memory cost of attention layers. For consistency, we used this model for all figures in the main text; however, in applications that do not require such light-weight networks, the larger models may be suitable.

\begin{table*}
    \begin{tabular}{c|c|c|c|c|c|c}
        Architecture & Block dimensions & Attention & Num. Parameters & \textbf{Mean FID}$\downarrow$ & \textbf{Median FID}$\downarrow$ & \textbf{Steps per second}$\uparrow$ \\
        \hline
        \rowcolor{lightgray}
        \texttt{Model-XS} & (32, 64, 128) & \xmark & 5.1M & 20.9 & 13.1 & \textbf{79.5/s} \\
        
        \rowcolor{white}
        \texttt{Model-S} & (32, 64, 64, 128) & $\mathtt{heads} = 4$, $\mathtt{dim}=32$ & 6.5M & 14.0 & 10.8 & 28.6/s \\
        
        \rowcolor{lightgray}
        \texttt{Model-M} & (64, 128, 128, 256) & $\mathtt{heads} = 4$, $\mathtt{dim}=32$ & 22.5M & \textbf{10.3} & \textbf{10.4} & 20.1/s
    \end{tabular}
    \caption{Three model architectures with varying parameter counts and design choices are shown. The \emph{Block dimensions} column shows the dimension counts for the downsampling blocks followed by the bottleneck block (the networks are symmetric; i.e. the upsampling blocks follow the same dimensions in reverse order). Our primary model (first row) is significantly faster than the other models, but compromises slightly on FID.}
    \label{tab:modelArchs}
\end{table*}

\begin{figure*}[p]
    \centering
    \includegraphics[width=\textwidth]{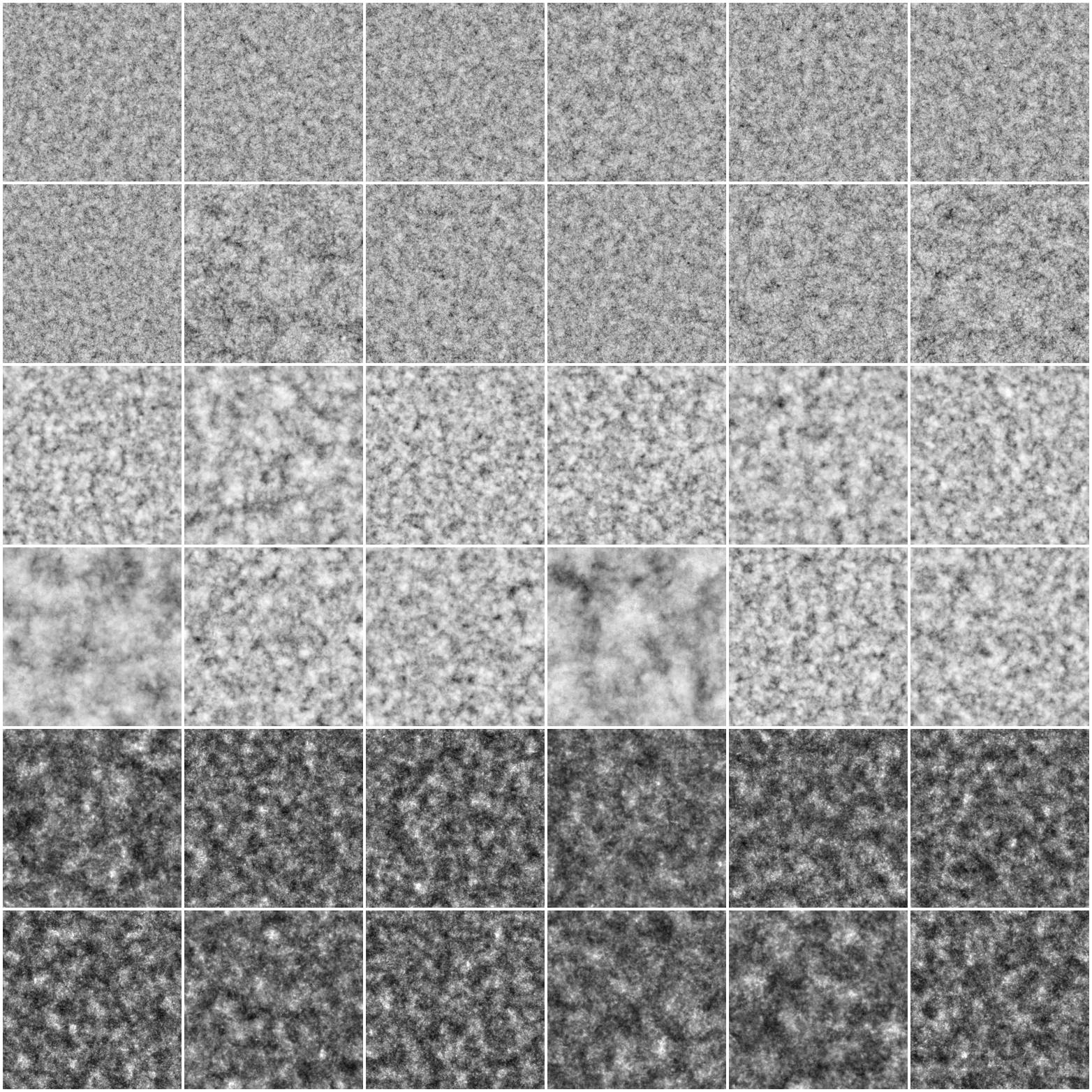}
    \caption{Random samples of our model's \texttt{clouds 1}, \texttt{clouds 2}, and \texttt{clouds 3} noises at $256\times256$ resolution.}
    \label{fig:grid5}
\end{figure*}

\end{document}